\documentclass[amsmath,amssymb,aps,prb,twocolumn,groupedaddress,showpacs,floatfix]{revtex4}
\usepackage{graphicx}

\begin{document}

\title{Quantum Fluctuations and Excitations in Antiferromagnetic Quasicrystals}
\author{Stefan Wessel$^{(1)}$} 
\author{Igor Milat$^{(2)}$}
\affiliation{$^{(1)}$Institut f\"ur Theoretische Physik III, Universit\"at Stuttgart, 70550 Stuttgart, Germany}
\affiliation{$^{(2)}$Theoretische Physik, ETH Z\"urich, CH-8093 Z\"urich, Switzerland}

\date{\today}
\begin{abstract}
We study the effects of quantum fluctuations and the excitation spectrum for the antiferromagnetic Heisenberg model on 
a two-dimensional
quasicrystal, by numerically solving linear  spin-wave theory on finite approximants of the octagonal tiling. 
Previous quantum Monte Carlo results for the distribution of local staggered magnetic moments and the 
static spin structure factor are reproduced well within this approximate scheme.
Furthermore, the magnetic excitation spectrum consists of 
magnon-like low-energy modes, as well as dispersionless high-energy states of multifractal nature. 
The dynamical spin structure factor, accessible to inelastic neutron scattering, 
exhibits linear-soft modes at low energies, self-similar structures with bifurcations emerging at intermediate 
energies,
and
flat bands in high-energy regions.
We find that the distribution of local staggered moments stemming from  the inhomogeneity of the quasiperiodic structure
leads to a characteristic energy spread in the local dynamical spin susceptibility, 
implying distinct nuclear magnetic resonance spectra, specific for 
different local environments.
\end{abstract}


\maketitle


\section{Introduction}
\label{sec:intr}
Quantum fluctuations are responsible for various degrees of disorder in low-dimensional quantum 
antiferromagnets. In particular two-dimensional systems show a variety of quantum
disordered phases, competing with conventional long-range magnetic order. 
For example, while the Heisenberg model on the square lattice 
exhibits true long-range order at zero 
temperature~\cite{manousakis}, spatial inhomogeneous magnetic exchange 
eventually leads to a complete suppression of magnetic order in structures 
such as the plaquette lattice, driven by  local singlet formation~\cite{plaquette,laeuchli}.
Other sources of magnetic disorder are frustration effects due to competing interactions~\cite{misguich}, and site/bond 
depletion~\cite{depletion}, where the reduction in the 
long-range magnetic 
order is accomplished  by  proliferation of localized low-energy excitations~\cite{mucciolo}.
All these systems share the translational invariance of the underlying lattice structure - assuming sufficient
self averaging in the case of quenched disorder. 

Quasiperiodic systems, lacking translational symmetry in addition to their inhomogeneous lattice structure were initially thought 
to not support sizeable correlations of localized magnetic moments.
However, recent neutron scattering 
experiments on  Zn-Hg-Ho~\cite{sato} as well as Cd-Mg-Tb~\cite{cdmgtb} icosahedral quasicrystals support the presence of 
significant magnetic 
correlations in these three-dimensional quasicrystalline compounds, with the absence of true long-range magnetic 
order~\cite{absence}  due to 
large frustrations in the 
antiferromagnetic exchange~\cite{wessel}.

\begin{figure}[t]
\begin{center}
\includegraphics[width=6cm]{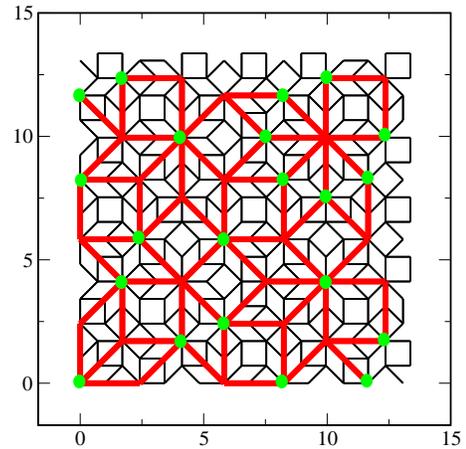}
\caption{
Finite approximant of the octagonal tiling with 239 sites (thin lines), along with a superimposed inflated 41 sites 
approximant (thick lines) of rescaled edge length by a factor of $\lambda=1+\sqrt{2}$. For the 41 sites approximant the vertices of one 
of the two sublattices are dressed with a disk to exhibit the bipartite nature of the octagonal tiling.
}
\label{fig:inflat}
\end{center}
\end{figure}

Bipartite, and thus unfrustrated, quasiperiodic crystal structures
were indeed shown to allow for sizeable two-sublattice antiferromagnetic order in a recent quantum Monte Carlo study of the spin-1/2 
Heisenberg 
model on the 
octagonal tiling~\cite{wessel}. Furthermore, the magnetic order in this two-dimensional system was found to exhibit
nontrivial patterns in 
the local staggered moment distribution, reflecting the self-similarity of the underlying quasiperiodic 
lattice structure~\cite{wessel}.
A  renormalization group approach based on this self-similarity indeed gives a gross account on the observed spread in the
staggered magnetization~\cite{anu}.
Furthermore, the static spin structure factor exhibits magnetic selection rules that impose a shift of reciprocal space 
indices~\cite{wessel}. 
The resulting neutron diffraction pattern~\cite{wessel} can be accounted for by analysis~\cite{lifshitz} of the quasicrystal 
spin 
group~\cite{ron}, also applicable to frustrated classical  models on the octagonal tiling with long-range exchange 
interactions~\cite{vedmedenko}.

While static properties of the Heisenberg antiferromagnet on the octagonal tiling are thus well studied, little is known about the 
spectral properties of these systems. On general grounds one would expect gapless Goldstone-modes to dominate 
at low-energies, even though the translational symmetry is absent. In addition, one would expect to find multifractional eigenstates,
as observed in tight-binding models on quasiperiodic lattices~\cite{grimm}.
Here, we investigate dynamical properties of quantum magnetic quasicrystals in order to 
identify the relevant energy scales of quantum fluctuations in such systems, determine the magnetic excitation spectrum, as 
well as to provide theoretical grounds for 
future experiments on magnetic quasicrystals, such as inelastic neutron scattering or magnetic resonance. In particular, 
we use linear spin-wave theory, which was successfully used in studies of periodic magnetically ordered systems, 
and apply it to the quasiperiodic case. 

The outline of the paper is as follows: Basic properties of the octagonal tiling are presented in the following section.  In 
Sec.~\ref{sec:line} we review linear spin wave theory in a real space formulation, and present
a numerical construction of the eigenmode expansion. An alternative scheme is given in the 
Appendix.
The results of applying this method to the octagonal tiling are discussed in Sec.~\ref{sec:resu}:
In Sec.~\ref{sec:stag} we discuss static 
properties of 
the magnetic correlations on the octagonal tiling, and compare our results to previous quantum Monte Carlo simulations. A detailed 
analysis of the excitation 
spectrum is presented in Sec.~\ref{sec:exci}, followed by a discussion of dynamical magnetic properties, such as the dynamical
spin structure factor in Sec.~\ref{sec:dyna}, and the local dynamical spin susceptibility (Sec.~\ref{sec:loca}). Finally, we conclude in 
Sec.~\ref{sec:conc} with a 
perspective on future investigations.

\section{Octagonal tiling}
\label{sec:octa}
In the following, we analyze the magnetic ground state properties and excitation spectrum of the nearest-neighbor 
antiferromagnetic spin$-1/2$ Heisenberg model,
\begin{equation}
 \label{eq:H-spin}
 H=J \sum_{\langle i,j \rangle} {\mathbf S}_i \cdot {\mathbf S}_j,\quad J>0, 
\end{equation}
on the most prominent example of a magnetic quasicrystal in two dimensions, the octagonal tiling.
The octagonal tiling is a bipartite quasiperiodic crystal system, and possesses an overall eightfold rotational symmetry, allowing for 
simple two-sublattice antiferromagnetism~\cite{lifshitz}. Sites in this tiling have coordination numbers $z$ ranging 
from 3 to 8, leading to a broad distribution of local staggered moments in the magnetically ordered ground state~\cite{wessel}.
A further important property of the octagonal tiling, in the absence of translational invariance, is its self-similarity under 
inflation 
transformations~\cite{levine}. This reversible operation refers to a well-defined decimation of a subset of vertices of the 
tiling, followed by a re-connection of the new vertices. Aside from a trivial rescaling of the length scale by a factor 
$\lambda=1+\sqrt{2}$, the infinite quasicrystal is left unchanged by this transformation. 

For our numerical study we consider finite square approximants of the octagonal tiling with $41$, $239$, and $1393$ sites.
These approximants can be obtained 
by the "cut-and-project" method
from a four-dimensional cubic lattice~\cite{duneau}, and 
are related by the inflation transformation, an example of which is shown in Fig.~\ref{fig:inflat}.
In order to avoid boundary-induced 
frustration effects, we 
apply toroidal boundary conditions~\cite{schulz}. 
Due to the lack of translational symmetry, we thus need to solve real-space linear spin-wave 
theory on lattices with up to 5572 sites. Before 
presenting our results, we provide details about the numerical scheme used in our calculations in the following section.

\section{Numerical spin-wave approximation}
\label{sec:line}
In this section, we review linear spin-wave theory, applied to the antiferromagnetic Heisenberg model on finite, bipartite 
lattices.
We consider a bipartite lattice with sublattices $A$ and $B$ consisting of $N_A$ and $N_B$ sites, respectively.
For the octagonal approximants considered in this work $N_A=N_B$, however
the following approach also applies if 
$N_A$ and $N_B$ are different. Following the standard Holstein-Primakoff approach\cite{spinwave}, we represent the spins in 
terms of 
bosonic operators $a_i$ and $b_i$. For $i\in A$, 
\begin{eqnarray}
  \label{eq:hpA}
  S^z_i & = &S - a^\dag_i a_i, \nonumber\\
  S^+_i &= &\sqrt{2S} \left(1 - \frac{a^\dag_i a_i}{2S}\right)^{1/2} a_i, \\
  S^-_i &= &\sqrt{2S} a^\dag_i \left(1 - \frac{a^\dag_i a_i}{2S}\right)^{1/2}, \nonumber
\end{eqnarray}
where $S$ denotes the spin magnitude, and
\begin{eqnarray}
  \label{eq:hpB}
  S^z_j & = &-S + b^\dag_j b_j, \nonumber\\
  S^+_j &= &\sqrt{2S} b^\dag_j \left( 1 - \frac{b^\dag_j b_j}{2S}\right)^{1/2},\\
  S^-_j &= &\sqrt{2S} \left(1 - \frac{b^\dag_j b_j}{2S}\right)^{1/2} b_j, \nonumber
\end{eqnarray}
for $j\in B$.
The linear spin-wave Hamiltonian is obtained by substituting the above identities in Eq.~(\ref{eq:H-spin}), 
expanding the square roots in of $1/S$, and keeping terms of lowest order $(1/S)^0$,
\begin{eqnarray}
  \label{eq:Hsw}
  H_{SW} &=& -J S(S + 1) N_b+ J S H_2,\\
  H_2&=& \sum_{\langle i,j \rangle}
  \left( a^\dag_i a_i + b_j b^\dag_j + a^\dag_i b^\dag_j + b_j a_i \right),
\end{eqnarray}
where $N_b$ denotes the number of bonds, and the sum in $H_2$ extends over all bonds of the bipartite lattice.
Introducing the $N_s=N_A+N_B$ component row vector
\begin{equation}
 \bar{a}^\dag=(a^\dag_1,...,a^\dag_{N_A},b_1,...,b_{N_B}),
\end{equation}
and the corresponding columnar conjugate, we can express the quadratic part of the Hamiltonian in matrix notation,
\begin{equation}
 H_2= \bar{a}^\dag M \bar{a} = \sum_{\langle i,j \rangle} \bar{a}^\dag M^{(i,j)} \bar{a},
\end{equation}
where
\begin{eqnarray}
 \label{eq:Mijkl}
 \left(M^{(i,j)}\right)_{k,l}&=&\delta_{i,k}\delta_{i,l}+\delta_{j+N_A,k}\delta_{j+N_A,l}\nonumber\\
                             & &+\delta_{i,k}\delta_{j+N_A,l}+\delta_{j+N_A,k}\delta_{i,l}
\end{eqnarray}
is the connectivity matrix of the lattice structure.
The bipartiteness of the lattice thus allows for a direct formulation in terms of the  $N_s \times N_s$ hermitian matrix $M$, 
instead of the 
general formulation
based on a $2 N_s \times 2 N_s$ matrix~\cite{blaizot}. 
We now seek a Bogoliubov transformation to the 
$N_n$ normal bosonic modes $\beta_k$, 
such that the spin-wave Hamiltonian $H_{SW}$ is diagonal when expressed in terms of the $\beta_k$,
\begin{equation}
  \label{eq:Hsw_diag}
  H_{SW} =N_s E_0+ \sum_{k=1}^{N_n} \omega_k \beta^\dag_k \beta_k,
\end{equation}
where $\omega_k>0$ denotes the eigenfrequency of the $k$-th mode, and $E_0$ the ground state energy per site.
To this end we make an Ansatz,
\begin{equation}
 \label{eq:T}
 \bar{a}=T\bar{\beta},
\end{equation}
with
\begin{equation}
 \bar{\beta}^\dag=(\beta^\dag_1,...,\beta^\dag_{N_+},\beta_{N_++1},...,\beta_{N_n}),
\end{equation}
so that we  divide the $N_n$ normal bosonic modes into two disjunct sets of
length $N_+$, and $N_-=N_n-N_+$, respectively, further determined
below.  Due to the bosonic commutation relations, the transformation
$T$ has to fulfill
\begin{equation}
 \label{eq:bcr}
 T \Gamma T^\dag = \Sigma,
\end{equation}
with matrices $\Gamma$ and $\Sigma$, defined by
\begin{equation} 
\label{eq:SL}
 \Gamma= \begin{pmatrix} 1_{N_+} & 0 \\ 0 & -1_{N_-} \end{pmatrix},
 \quad\Sigma= \begin{pmatrix} 1_{N_A} & 0 \\ 0 & -1_{N_B} \end{pmatrix}.
\end{equation}
Here, $1_{N}$ denotes the $N\times N$ identity matrix~\cite{white}.
Since $\Sigma^2 = 1$, as well as $\Gamma^2 = 1$, 
and because the left and right
inverse of a square matrix are identical, 
if $T$  satisfies
\begin{equation}
\label{eq:constr}
T^\dag \Sigma T = \Gamma,
\end{equation}
the bosonic commutation relations are fulfilled. Since $T$
diagonalizes $H_2$, we have
\begin{equation}
 T^\dag M T= \Omega, \quad \Omega={\rm diag}(\omega_1,...,\omega_{N_n}),
\end{equation}
from which we find upon multiplication from the left with $T \Gamma$,
that $T$ has to satisfy
\begin{equation}
 \Sigma M T=T\Gamma \Omega.
\end{equation}
The column vectors of $T$ are thus seen to be related to the right
eigenvectors of $\Sigma M$.  For semi-positive $M$, $\Sigma M$ has
real eigenvalues, and eigenvectors belonging to different eigenvalues
are orthogonal~\cite{blaizot}.  We denote the number of positive
(negative) eigenvalues by $N_+$ ($N_-$), the number of zero-modes by
$N_0$, and label the positive (negative) eigenvalues by $\lambda^+_i$,
$i=1,...,N_+$ ($\lambda^-_i$, $i=1,...,N_-$).  After numerically
solving the non-hermitian $N_s\times N_s$ eigenvalue problem for
$\Sigma M$~\cite{lapack}, we construct within the subspace of each
degenerate eigenvalue $\lambda^{\pm}_n>0 (<0)$ of dimension
$d^{\pm}_n$ eigenvectors $z^{\pm}_{n,1},...,z^{\pm}_{m,d^{\pm}_n}$,
obeying
\begin{equation}
 \label{eq:orthonormal}
 (z^{\pm}_{n,i})^\dag\Sigma z^{\pm}_{n,j}=\pm \delta_{i,j},\quad i,j=1,...,d^{\pm}_n\
\end{equation}
using Gram-Schmidt orthogonalization with respect to
$\Sigma$ (ref.~\onlinecite{lang}).  We then obtain $T$ from the orthonormal
eigenvectors as
\begin{equation}
 \label{eq:Tmatrix}
 T=(z^+_{1,1},...,z^+_{N_+,d^+_{N_+}},z^-_{1,d_1},...,z^-_{N_-,d^-_{N_-}}),
\end{equation}
where Eq.~(\ref{eq:orthonormal}) ensures that Eq.~(\ref{eq:constr}) is
satisfied.  The corresponding eigenfrequencies are given by
\begin{eqnarray}
\omega_k&=&J S \lambda^+_i,\quad k=1,...,N_+,\\
\omega_{N_++k}&=&- J S \lambda^-_k,\quad k=1,...,N_-.
\end{eqnarray}
The zero-modes of $\Sigma M$ correspond to collective modes due to the
broken continuous symmetry implied by the classical N\'eel
state~\cite{blaizot,mucciolo}.  Furthermore, the ground state energy becomes
\begin{equation}
E_0 N_s =-J S(S + 1) N_b+ J S \sum_{k=1}^{N_-} |\lambda^{-}_k|.
\end{equation}
An alternative means of numerically constructing the transformation matrix $T$ is presented in the appendix. We have 
verified, that both approaches indeed yield the same results.

\section{Results}
\label{sec:resu}
In this section, we present the results obtained by applying the linear spin-wave approximation to the octagonal tiling
introduced in Sec.~\ref{sec:octa}.
Within linear spin-wave theory, ground state expectation values of both static and dynamic magnetic correlations can 
be 
calculated from contractions of bosonic normal mode operators. It turns out convenient for this purpose, to
express the Bogoliubov transformation in terms of the row vectors
\begin{eqnarray}
 \tilde{a}^\dag&=&(a^\dag_1,...,a^\dag_{N_A},b^\dag_1,...,b^\dag_{N_B},a_1,...,a_{N_A},b_1,...,b_{N_B}),\nonumber\\
 \tilde{\beta}^\dag&=&(\beta^\dag_1,...,\beta^\dag_{N_n},\beta_1,...,\beta_{N_n}),\nonumber
\end{eqnarray}
as
\begin{equation}
 \label{eq:tildeT}
 \tilde{a}=\tilde{T}\tilde{\beta}, \quad \tilde{T}=\begin{pmatrix} U & V \\ V^* & U^* \end{pmatrix}.
\end{equation}
One obtains $U$ and $V$ from the transformation matrix $T$ of Eq.~(\ref{eq:T}), upon
defining 
$N_+\times N_+$ matrices $A$ and $B$, and $N_- \times N_-$ matrices $C$ and $D$ such that
\begin{equation}
 T=\begin{pmatrix} A & C \\ B & D \end{pmatrix},
\end{equation}
and gets
\begin{equation}
 U=\begin{pmatrix} A & 0 \\ 0 & D^* \end{pmatrix}, \quad V=\begin{pmatrix} 0 & C \\ B^* & 0 \end{pmatrix}. 
\end{equation}

\subsection{Staggered magnetization}
\label{sec:stag}
\begin{figure}[t]
\begin{center}
\includegraphics[width=8cm]{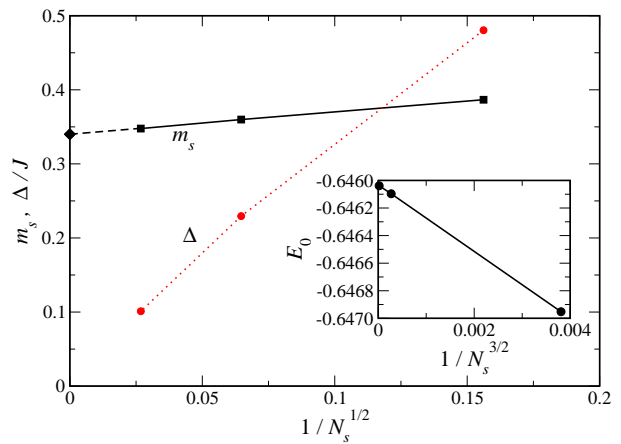}
\caption{
Finite size scaling of the ground state staggered magnetization, $m_s$, and the lowest excitation gap, $\Delta$, for
the $S=1/2$ Heisenberg model on the octagonal tiling in linear spin-wave theory. The inset shows the finite size scaling of
the ground state energy, $E_0$, for the same system.
}
\label{fig:fss}
\end{center}
\end{figure}

In this section we examine static properties of the magnetic ground state in the octagonal tiling.
For this purpose, we first calculate the staggered magnetization at each lattice site $i$
in linear spin-wave theory,
\begin{equation}
 \label{eq:lsm}
 m_s(i)=|\langle S^z_i \rangle |=S- \sum_k |V_{ik}|^2. 
\end{equation}
The spatially averaged staggered magnetization 
\begin{equation}
m_s=\sum_{i=1}^{N_s} m_s(i)
\end{equation}
is shown as a function of the system size $N_s$ of the approximant of the octagonal tiling
in Fig.~\ref{fig:fss}. The finite size values scale well as a function of $N_s^{-1/2}$, and indicate a sizeable staggered 
magnetization of the long ranged ordered ground state in the octagonal tiling.
The value of $m_s$ extrapolated to the thermodynamic limit is $m_s=0.34$,
and agrees well with the quantum Monte Carlo result, ${m}_s=0.337\pm 0.002$~\cite{wessel}. 
The  weak reduction of the order parameter by quantum fluctuations is indicative for the feasibility  of the linear 
spin-wave approach in this system. The ground state energy $E_0$, shown in the inset of Fig.~\ref{fig:fss}, is also found to 
scale well as a function of $N_s^{-3/2}$, as 
expected for an ordered state~\cite{einarsson}. The extrapolated value in the thermodynamic limit, $E_0=-0.646$, compares well 
with the quantum Monte Carlo result, $E_0=-0.6581(1)$~\cite{wessel_e0}.

\begin{figure}[t]
\begin{center}
\includegraphics[width=8cm]{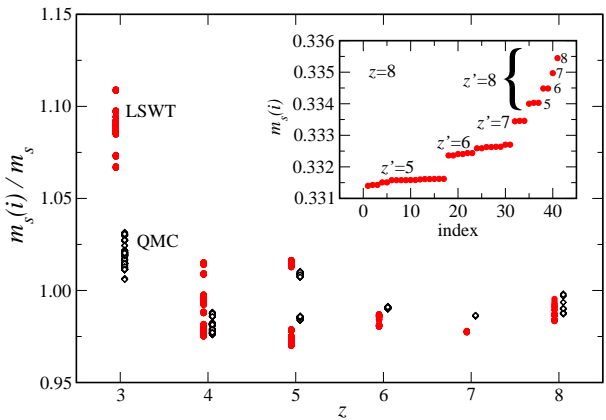}
\caption{
Dependence of the local staggered magnetization on the coordination number $z$ for all sites in the 1393 sites 
approximant of the octagonal tiling within linear spin-wave theory of the $S=1/2$ Heisenberg model (LSWT). For reference,  results 
of quantum Monte Carlo simulations of the same system (QMC) are also shown~\cite{wessel}. The inset exhibits the hierarchy of 
the 
local 
staggered magnetization of the $z=8$ sites, grouped according to the value of $z'$ under a deflation transformation. Numbers next
to symbols give the value of $z''$ for $z'=8$ sites under a further deflation.
}
\label{fig:lsm}
\end{center}
\end{figure}

In Ref.~\onlinecite{wessel}, the magnetic ground state on the octagonal tiling was found to exhibit a nontrivial 
local structure reflecting the self-similarity of the underlying quasiperiodic lattice structure. 
We now analyze to what extent spin-wave theory is able to reproduce this structure and thus to account for 
the specific nature of quantum fluctuations in an inhomogeneous connectivity.
For this purpose, we show in Fig.~\ref{fig:lsm} the linear spin-wave results of the local staggered magnetization, 
$m_s(i)$, for the 1393 sites approximant, and compare those with values grouped according to the coordination number $z$ of the various 
sites,
with the quantum Monte Carlo data of Ref.~\onlinecite{wessel}. 
  
We find linear spin-wave theory to qualitatively reproduce characteristic features of the local staggered moment distribution, 
such as (i) a wide 
spread of the moments, in particular for small values of $z$, (ii) a prominent bimodal splitting of the moments for $z=5$ and (iii) 
the hierarchical structure observed in the splitting of the moments for sites with $z=8$, shown in the inset of Fig.~\ref{fig:lsm}. 
These splittings in the local staggered moments can be accounted for by the 
properties of inequivalent sites under deflation transformations, reflecting their different local 
environments~\cite{wessel}.
In particular, fivefold sites with $z=5$ always occur in pairs, with two different types of  site. The first type is connected to 
four 
fourfold ($z=4$)
sites and the other fivefold site, while for the other type two neighbors are fourfold, two threefold ($z=3$) and one 
fivefold.
This difference in the local connectivity leads to the observed splitting~\cite{wessel}, and 
is also reflected in a  different behavior of the two types of fivefold sites under 
deflation transformations: while one type is decimated, the other remains as a threefold site in the deflated tiling.
The eightfold ($z=8$) sites exhibit an even richer, hierarchical structure shown in the inset of Fig.~\ref{fig:lsm}, with
moments grouped  according to the different deflation properties of the eightfold sites, 
namely their new coordination number $z'$. The high symmetry sites with $z'=8$ show a further hyperfine splitting, with 
moments grouped according to the value of  $z''$ under a second deflation, indicated by numbers next to the symbols in the inset 
of Fig.~\ref{fig:lsm}. As discussed 
in Ref.~\onlinecite{wessel}, these splittings eventually
lead to a multifractal distribution of local staggered moments in the infinite quasicrystal,  for this class of sites.

From a quantitative comparison of the local staggered magnetization between linear spin-wave theory and quantum Monte Carlo, 
we find 
characteristic limitations of the spin-wave 
approach to persist on a local level. Namely, while local staggered moments of high-connectivity sites are 
reproduced even quantitatively, deviations of 
about $8\%$ are observed for low-connectivity sites. In an inhomogeneous environment, linear spin-wave theory is thus 
more accurate at sites of large coordination, as might have been expected from its behavior in homogenous systems. 

\begin{figure}[t]
\begin{center}
\includegraphics[width=4cm]{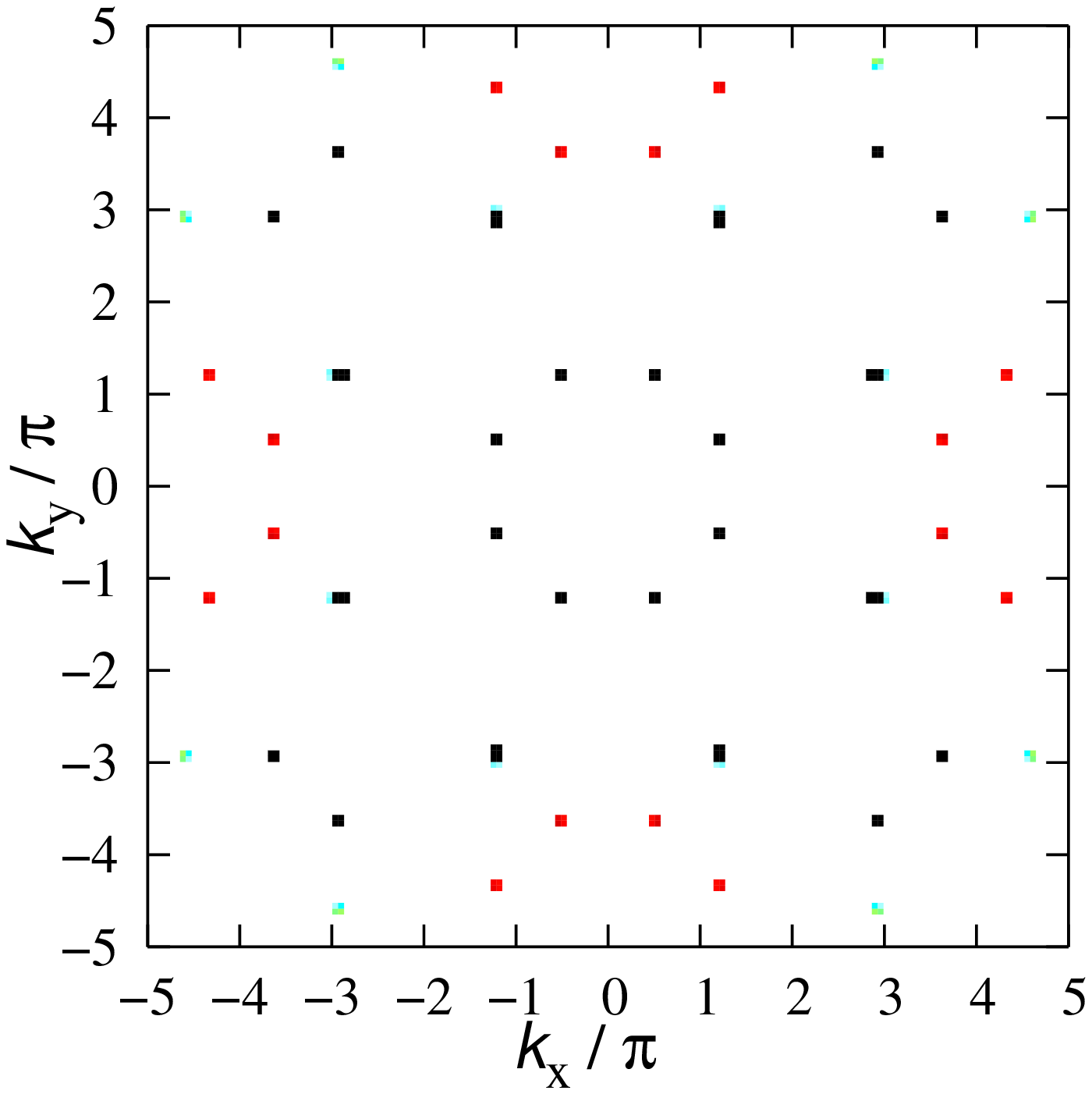}
\includegraphics[width=4cm]{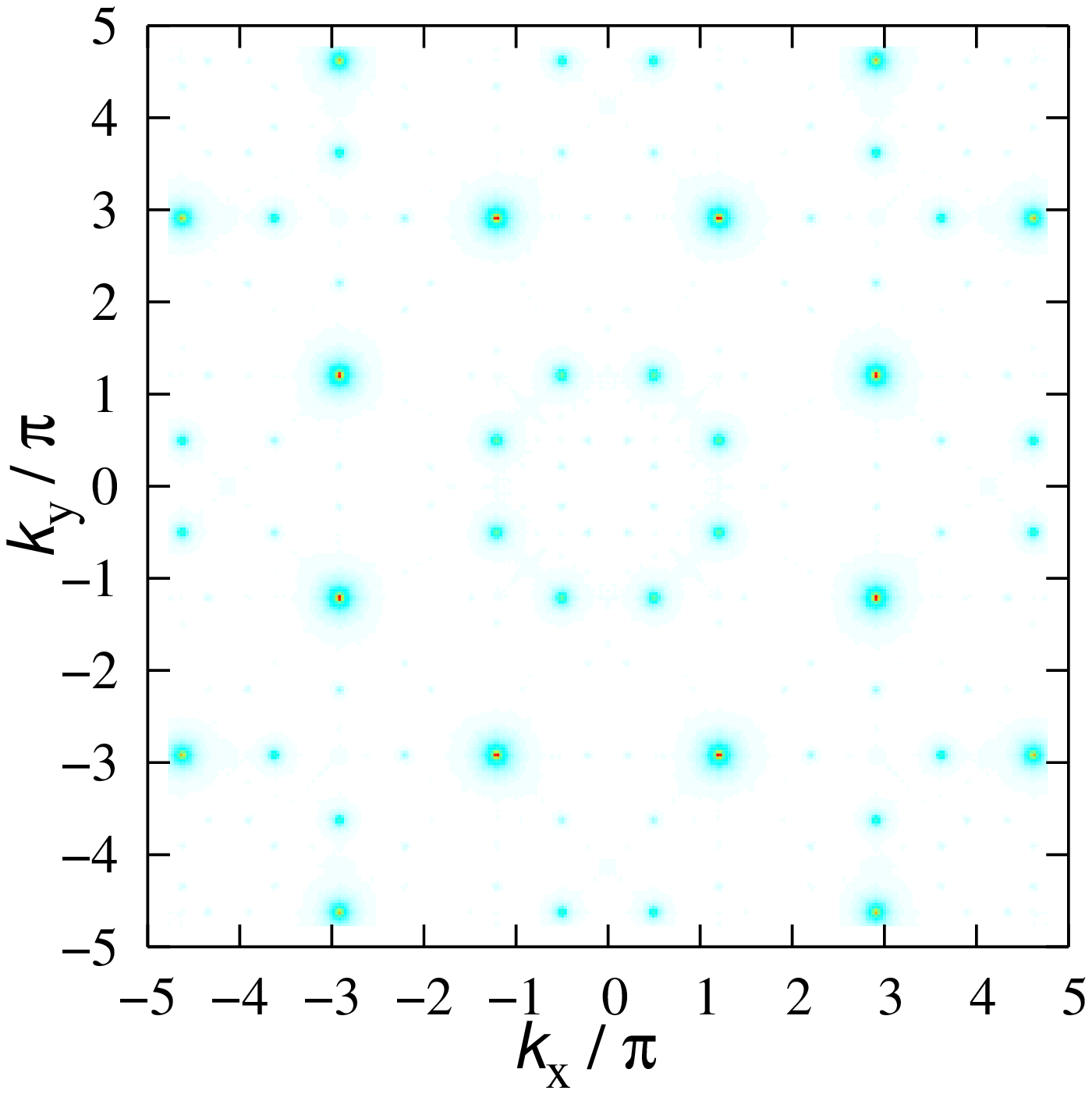}
\caption{
Intensity plot of the static longitudinal magnetic structure factor $S^{\parallel}({\mathbf k})$ (left), and the 
integrated dynamical 
spin structure factor $S^{\perp}({\mathbf k})$ (right)
for the $S=1/2$ Heisenberg antiferromagnet on the 1393 sites approximant of the 
octagonal tiling.
}
\label{fig:Sxx_Szk}
\end{center}
\end{figure}

The long-range antiferromagnetic order in the octagonal tiling
leads to characteristic neutron diffraction patterns,
due to selection rules imposed by the magnetic symmetry~\cite{wessel,lifshitz}.
These patterns can be obtained  from the 
static longitudinal structure factor 
\begin{equation}
S^{\parallel}({\mathbf k})=\sum_{i,j=1}^{N_s} e^{i{\mathbf k} \cdot ({\mathbf r}_i - {\mathbf r}_j)} \langle S^z_i S^z_j \rangle,
\end{equation}
which within linear spin-wave theory amounts to the Fourier transform of the real space distribution of $\langle S^z_i\rangle$.
Since linear spin-wave theory reproduces the spatial staggered moment distribution rather well, as seen from Fig.~\ref{fig:lsm}, 
the
resulting static longitudinal structure factor, shown in the left of Fig.~\ref{fig:Sxx_Szk} also compares well 
to the quantum Monte Carlo result~\cite{wessel}. In particular, it exhibits the extinction of 
nuclear Bragg peaks and the emergence of new, magnetic Bragg peaks. For a 
detailed theoretical derivation  of the various selection rules in the octagonal tiling, we refer to the explicit 
enumerations in Ref.~\onlinecite{lifshitz}.

\begin{figure}[t]
\begin{center}
\includegraphics[width=8cm]{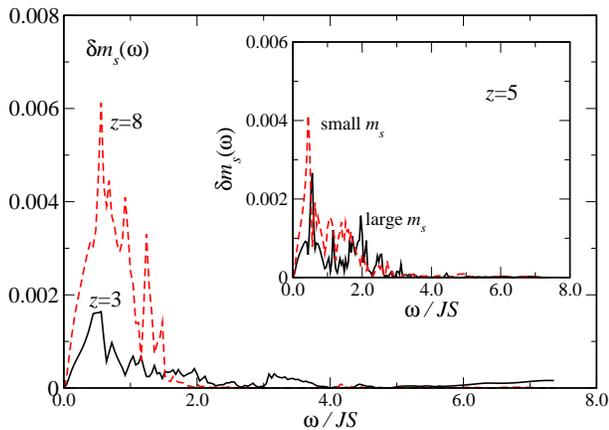}
\caption{
Frequency dependence of quantum fluctuations to the staggered magnetization, $\delta m_s(\omega)$, for the $S=1/2$ 
Heisenberg model on the octagonal tiling, averaged separately over sites with 
coordination numbers $z=3$ and $8$. The inset shows $\delta m_s(\omega)$ for $z=5$, averaged separately over sites with a
small and  large staggered magnetization, respectively.
}
\label{fig:dms}
\end{center}
\end{figure}

The structural distribution of local staggered moments observed in Fig.~\ref{fig:lsm} exhibits an inhomogeneous distribution of 
quantum
fluctuations. As seen from Eq.~(\ref{eq:lsm}), fluctuations in $m_s$ arise from distinct contributions from the various 
eigenstates of the
system. 
To quantify the relevance of the different eigenstates for the quantum fluctuations, 
we study the reduction of the staggered magnetization at each lattice site $i$, which can be parameterized as a function of 
energy,
\begin{equation}
\delta m_s(i,\omega)=\frac{\sum_k |V_{ik}|^2 \delta(\omega-\omega_k)}{\sum_k \delta(\omega-\omega_k)}.
\label{eq:dms}
\end{equation}
In Fig.~\ref{fig:dms} we show $\delta m_s(i,\omega)$, averaged separately over sites with $z=3$, and 
$z=8$.~\cite{footnote1}

In both cases, the dominant contribution to quantum fluctuations stem from the low-energy modes, with  $\omega/JS<2$. 
For threefold sites,
further contributions to $\delta m_s$  arise from higher energy modes, which are not relevant for the eightfold sites. 
In general, we find the  upper bound of the 
energy range that is relevant for quantum 
fluctuations  to
decrease with increasing coordination number $z$. 
Although sites with low coordination numbers thus receive quantum fluctuations over a larger range in energy space, their
staggered moment is typically larger than for high coordinated sites, as seen from Fig.~\ref{fig:lsm}.
A transfer of relevant quantum fluctuations
to low-energy modes is thus responsible for a decrease of the staggered moment at specific sites. 
This increased relevance of low-energy modes for quantum fluctuations 
is also observed for sites with the same coordination number. 
For example, the inset of 
Fig.~\ref{fig:dms} shows $\delta m_s$ for the fivefold sites, where the bimodal splitting was observed in Fig.~\ref{fig:lsm}, 
averaged 
separately over sites with a small and a large moment, respectively. 
The additional reduction of the local moment for one type of fivefold sites is clearly seen to be due to a proliferation of 
quantum fluctuations in the lower energy region.

This hints at a close  link between the structure of the inhomogeneous magnetic ground state and the
magnetic excitation spectrum.
In the following subsection, we proceed to analyze the magnetic excitations 
in more detail.

\subsection{Excitation Spectrum}
\label{sec:exci}
In the previous section we found that
spin-wave theory provides a qualitative, and even quantitative account on
the magnetic ground state properties of the Heisenberg model on the   
octagonal tiling. 
We now proceed to use linear spin-wave theory to gain insights into the 
spectral properties of this quasiperiodic antiferromagnet.

\begin{figure}[t]
\begin{center}
\includegraphics[width=8cm]{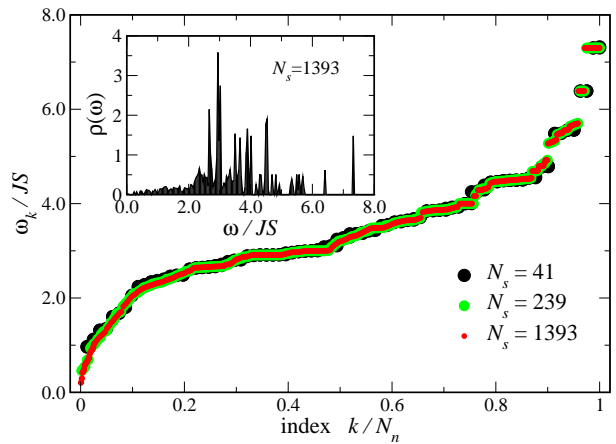}
\caption{
Excitation spectra of the Heisenberg antiferromagnet on finite approximants of the octagonal tiling of different sizes within 
linear spin-wave theory. The inset shows the density of state, $\rho(\omega)$, for the largest approximant. 
}
\label{fig:dos}
\end{center}
\end{figure}

The spectra of spin-wave normal modes for the first three approximants of the octagonal tiling
are shown in Fig.~\ref{fig:dos}. For finite approximants the spectra consists of discrete sets of 
energy levels, spanning an extended range up to $\omega_{max}/JS\approx 7.3$. This implies a bandwidth almost a factor two
larger than for the Heisenberg model on the square lattice, where $\omega_{max}/JS=4$.
Upon increasing the system size, the energy spectrum appears to develop a
dense band, and two  isolated pockets at higher energies, near $\omega/JS\approx6.5$ and $7.3$, respectively. The presence of 
a third 
pocket near $\omega/JS\approx5.5$ cannot be excluded from the finite size data. However, from Fig.~\ref{fig:dos} the gap near 
$\omega/JS\approx 5$ appears to eventually close for higher approximants, whereas 
the gaps to the  higher energy pockets do not show 
any decrease for the approximants considered here.

For a more quantitative analysis of the distribution of normal modes, we calculate the density of states of the spin-wave excitation 
spectrum (DOS), 
\begin{equation}
\rho(\omega)=\sum_k \delta(\omega-\omega_k),
\end{equation}
shown for the largest approximant in the inset of Fig.~\ref{fig:dos}. The spin-wave DOS exhibits a characteristic spiky shape in the 
high-energy regime, in particular for energies larger than 
$\omega/JS\approx 3$, similar to shapes found for the tight-binding 
Hamiltonian on the 
same lattice structure~\cite{anu2000}.
In the low-energy region, the DOS appears more smooth, with a residual 
roughness due to the limited resolution of the  
energy spectra due to finite size effects. We indeed find similar resolution limited roughness also on finite square lattice systems.

\begin{figure}[t]
\begin{center}
\includegraphics[width=8cm]{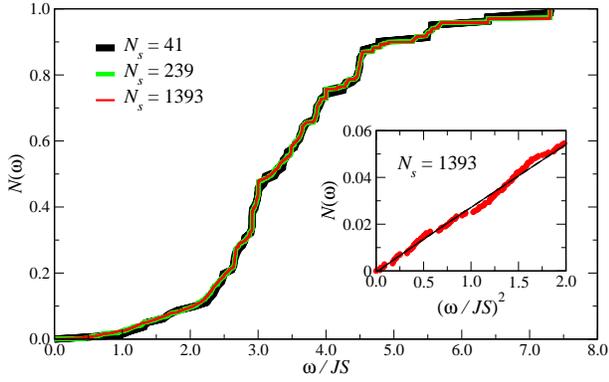}
\caption{
Cumulative density of states, $N(\omega)$, of the Heisenberg antiferromagnet on finite approximants of the octagonal tiling of 
different sizes within linear spin-wave theory. The inset exhibits the quadratic scaling of $N(\omega)$ at low energies 
for the largest approximant, indicative of a linear low-energy density of states.
}
\label{fig:cdos}
\end{center}
\end{figure}

On general grounds, the presence of a long-range ordered ground state is expected to characterise  the low-energy properties 
of 
the excitation spectrum.
In particular, due to the broken $SU(2)$ symmetry, we expect 
the gap to the lowest excitation, $\Delta$, to close
in the thermodynamic limit of the quasiperiodic tiling.
A finite size scaling analysis of $\Delta$, shown in 
Fig.~\ref{fig:fss}, is indeed consistent with $\Delta\propto N_s^{-\alpha}$, where $\alpha\approx 0.5$.
At low energies the DOS furthermore increases linearly, $\rho(\omega)\propto\omega$,
as seen from the quadratic low-energy behavior of the 
cumulative density of states,
\begin{equation}
N(\omega)=\int_0^{\omega} d\epsilon \: \rho(\epsilon),
\end{equation}
shown in Fig.~\ref{fig:cdos}. Here, we employ $N(\omega)$, since this quantity is less susceptible to resolution limited 
roughness than $\rho(\omega)$, due to the frequency integration.
The low-energy features of the spin-wave DOS in the octagonal tiling are also observed 
in  periodic antiferromagnets. For example, on the square lattice, antiferromagnetic 
spin-waves obey a linear dispersion relation at low energies
$\omega_{\mathbf k}\approx \sqrt{8}JS |{\mathbf k}|$,
resulting in a linear low-energy DOS in this system.


The similarity of the low-energy DOS for the Heisenberg model on the quasiperiodic octagonal tiling to the periodic case suggest 
that the low-energy excitation could be delocalized, magnon-like modes also in the octagonal tiling. We are thus lead to 
analyze the spatial extent of the 
eigenstates found within linear spin-wave theory.
An appropriate method to characterize the localization properties
is the inverse participation ratio (IPR)~\cite{wegner} which expresses the spatial extent of the wavefunction of a given state.
For the bosonic  excitations of quantum magnets, the IPR of an eigenstate is given 
by its contribution to the quantum fluctuations of the staggered magnetization.
Following Ref.~\onlinecite{mucciolo},
we  define the energy dependent inverse participation ratio 
\begin{equation}
 I(\omega)=\frac{\sum_k\: I_k\: \delta(\omega-\omega_k)}{\sum_k \delta(\omega-\omega_k)},
\end{equation}
where
\begin{equation}
 I_k=\frac{\sum_{i} |V_{ik}|^4}{\left(\sum_{i} |V_{ik}|^2 \right)^2},
\end{equation}
and study its scaling behavior upon increasing the system size. 
For delocalized states of a $d$ dimensional quantum system, the IPR 
decreases with the system size, $N_s$, as $N_s^{-1}$.
Exponentially localized states should be very insensitive to the system size and one expects a size independent IPR.

\begin{figure}[t]
\begin{center}
\includegraphics[width=8cm]{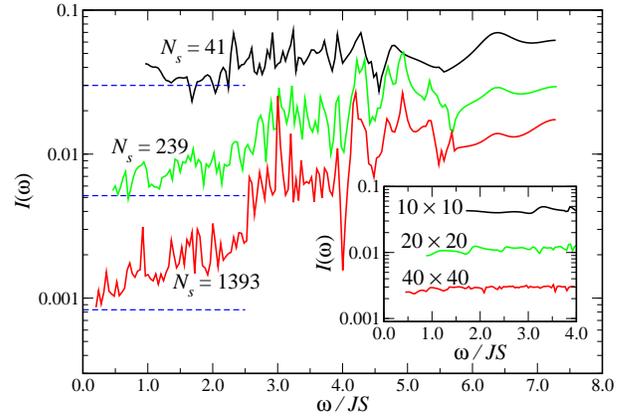}
\caption{
Frequency dependent inverse participation ratio, $I(\omega)$, for the $S=1/2$ Heisenberg antiferromagnet on finite approximants 
of the octagonal tiling of different sizes, obtained within linear spin-wave theory. Dashed lines indicate a finite size scaling 
as $N_s^{-1}$, expected for extended states.
The inset shows linear spin-wave results for $I(\omega)$ on square lattices of different sizes. 
}
\label{fig:ipr}
\end{center}
\end{figure}

Fig.~\ref{fig:ipr} shows the calculated IPR as a function of energy for the spin$-1/2$ Heisenberg model on finite approximants
of the octagonal tiling. For comparison, we show in the inset the corresponding quantity for the Heisenberg model 
on the square lattice. In  the square lattice case, all eigenstates show the characteristic $N_s^{-1}$ scaling, expected for
extended magnon states. For the octagonal tiling, we find such behavior only for the low-energy modes, but due to the limited 
resolution cannot exclude a reduced finite size scaling down to zero frequency.
The low-energy modes in the octagonal tiling thus appear as extended excitations out of the antiferromagnetic ground state, 
similar to coherent magnons in the periodic case. The higher energy states also do not appear exponential localized, but 
show a significantly reduced finite size scaling 
$I(\omega)\propto N_s^{-\beta}$, with a scaling exponent $\beta<1$, characteristic for multifractal states, as observed for 
critical states at the Anderson localization transition~\cite{schreiber}. 
The effective exponent $\beta$ decreases towards the upper edge of the spectrum, being lowest for energies near the
isolated pockets found in the DOS in Fig.~\ref{fig:dos}. 
While a reliable determination of $\beta$ would require the study of substantially larger approximants, we estimate a value of 
$\beta\approx 0.2$ for the high-energy modes.
The multifractality of eigenstates 
found in non-interacting models on quasiperiodic crystals~\cite{grimm} is thus clearly observed also for excitations of 
strongly correlated systems, such as the Heisenberg antiferromagnet.

\subsection{Dynamical spin structure factor}
\label{sec:dyna}
Having analyzed the spin-wave excitations of the Heisenberg antiferromagnet on the octagonal tiling,
we  proceed to study magnetic properties exhibiting the magnetic excitation spectrum, such as the dynamical spin structure 
factor, and  local dynamical spin susceptibility.

\begin{figure}[t]
\begin{center}
\includegraphics[width=4cm]{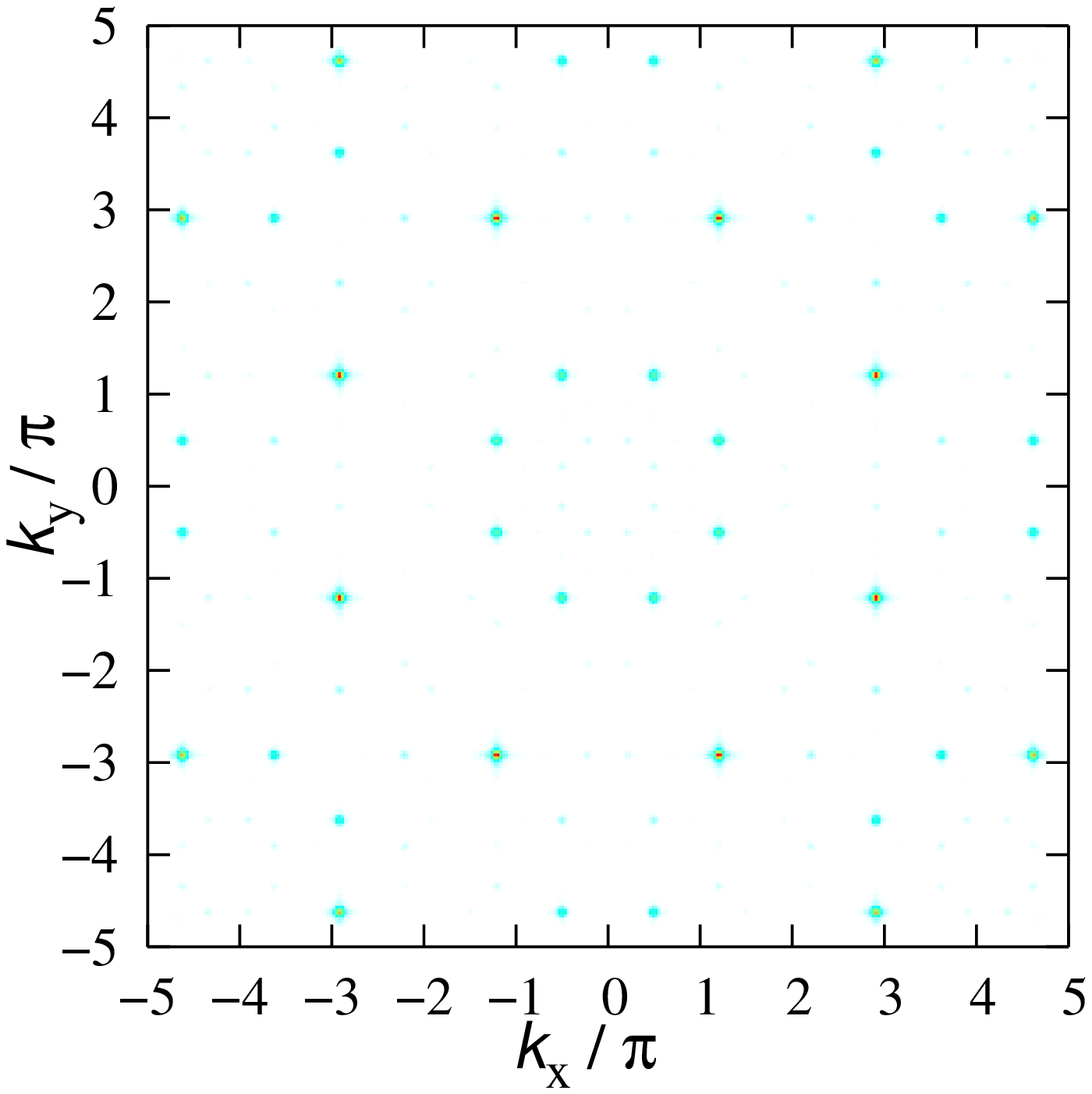}
\includegraphics[width=4cm]{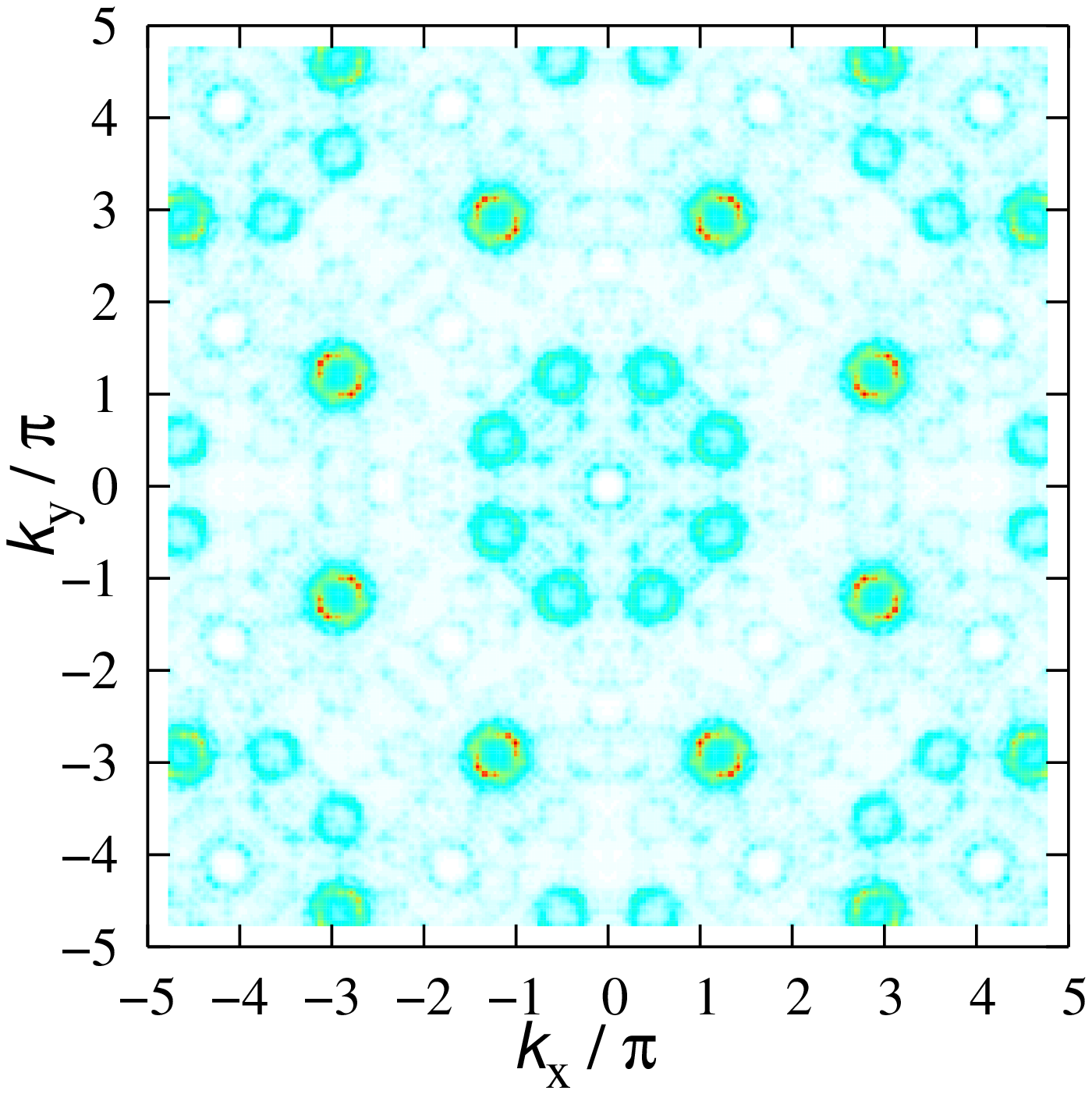}
\caption{
Intensity plot of the dynamical spin structure factor, $S({\mathbf k},\omega)$,
for fixed $\omega/JS=0$ (left) and $\omega/JS=1.8$ (right)
for the $S=1/2$ Heisenberg antiferromagnet on  the 1393 sites approximant of the 
octagonal tiling.
}
\label{fig:Sxx_o_1}
\end{center}
\end{figure}

\begin{figure}[t]
\begin{center}
\includegraphics[width=8cm]{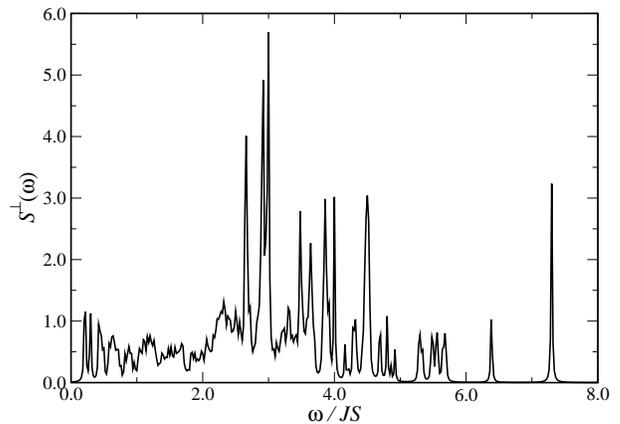}
\caption{
Frequency dependence of the momentum-integrated dynamical spin structure factor, $S^{\perp}(\omega)$, 
for the $S=1/2$ Heisenberg antiferromagnet on the octagonal tiling.
}
\label{fig:Sxx_o_all}
\end{center}
\end{figure}

\begin{figure}[t]
\begin{center}
\includegraphics[width=4cm]{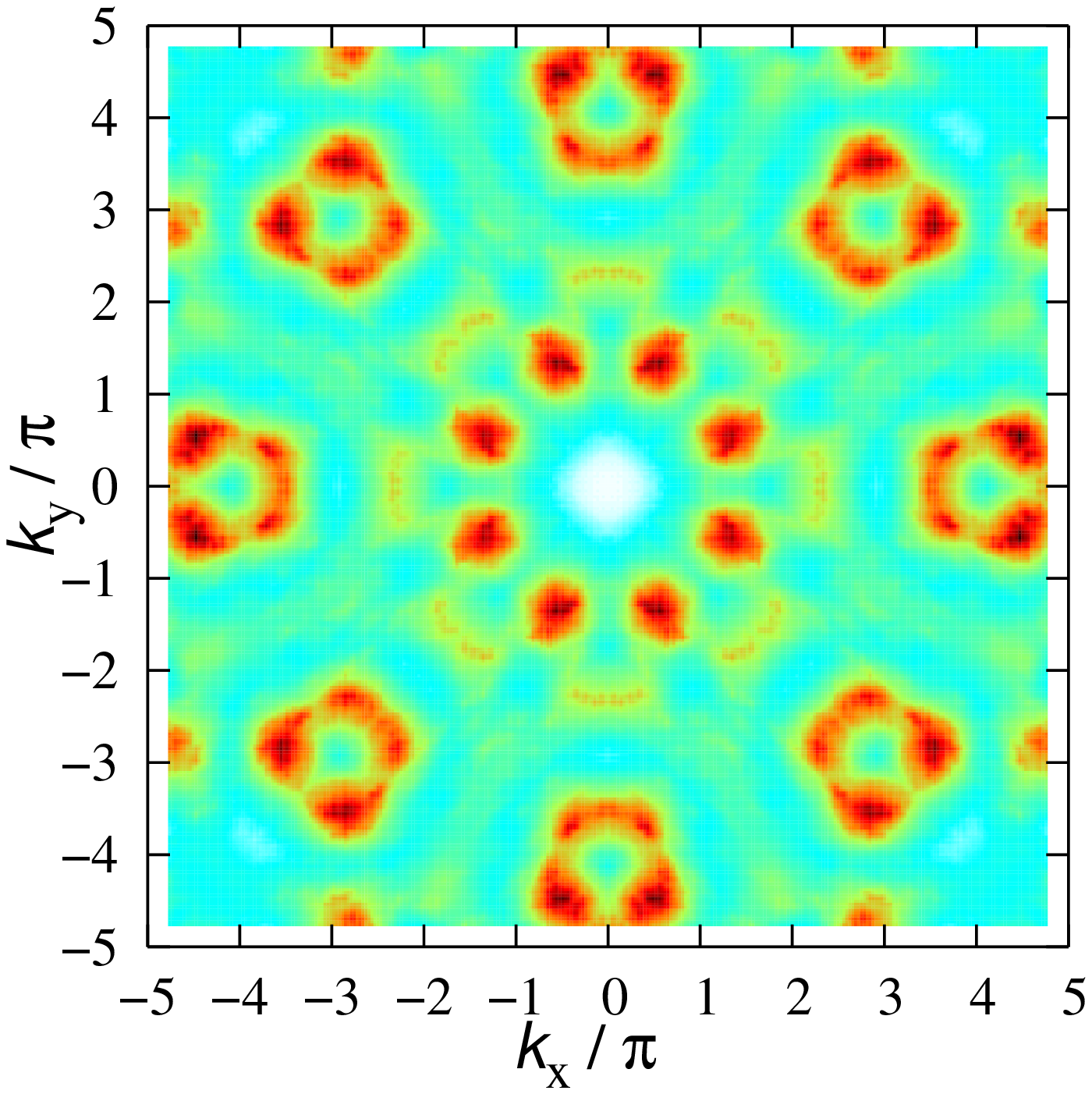}
\includegraphics[width=4cm]{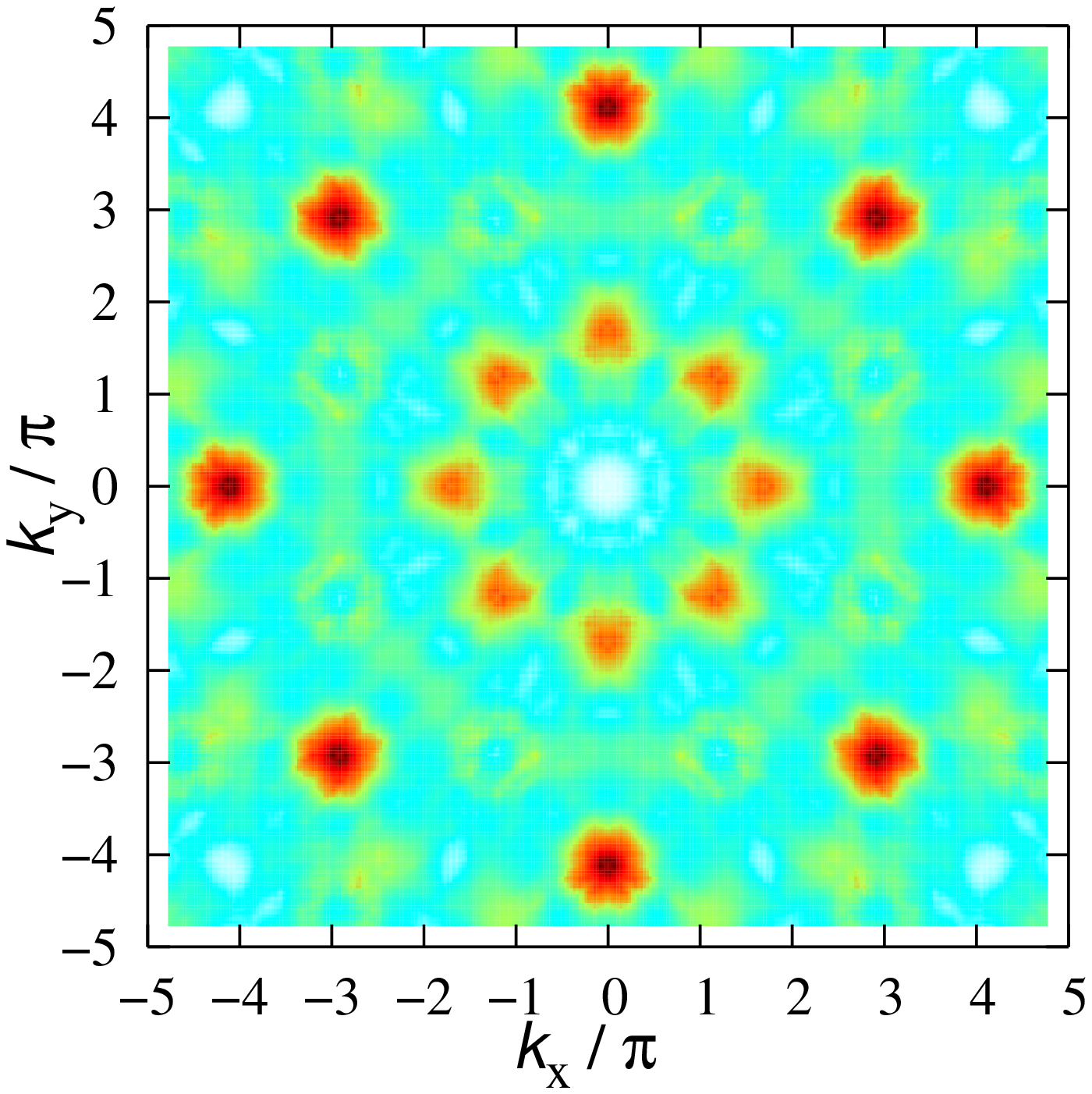}
\caption{
Intensity plot of the dynamical spin structure factor, $S({\mathbf k},\omega)$,
for fixed $\omega/JS=2.9$ (left) and $\omega/JS=3.0$ (right)
for the $S=1/2$ Heisenberg antiferromagnet on  the 1393 sites approximant of the 
octagonal tiling.
}
\label{fig:Sxx_o_2}
\end{center}
\end{figure}

\begin{figure}[t]
\begin{center}
\includegraphics[width=4cm]{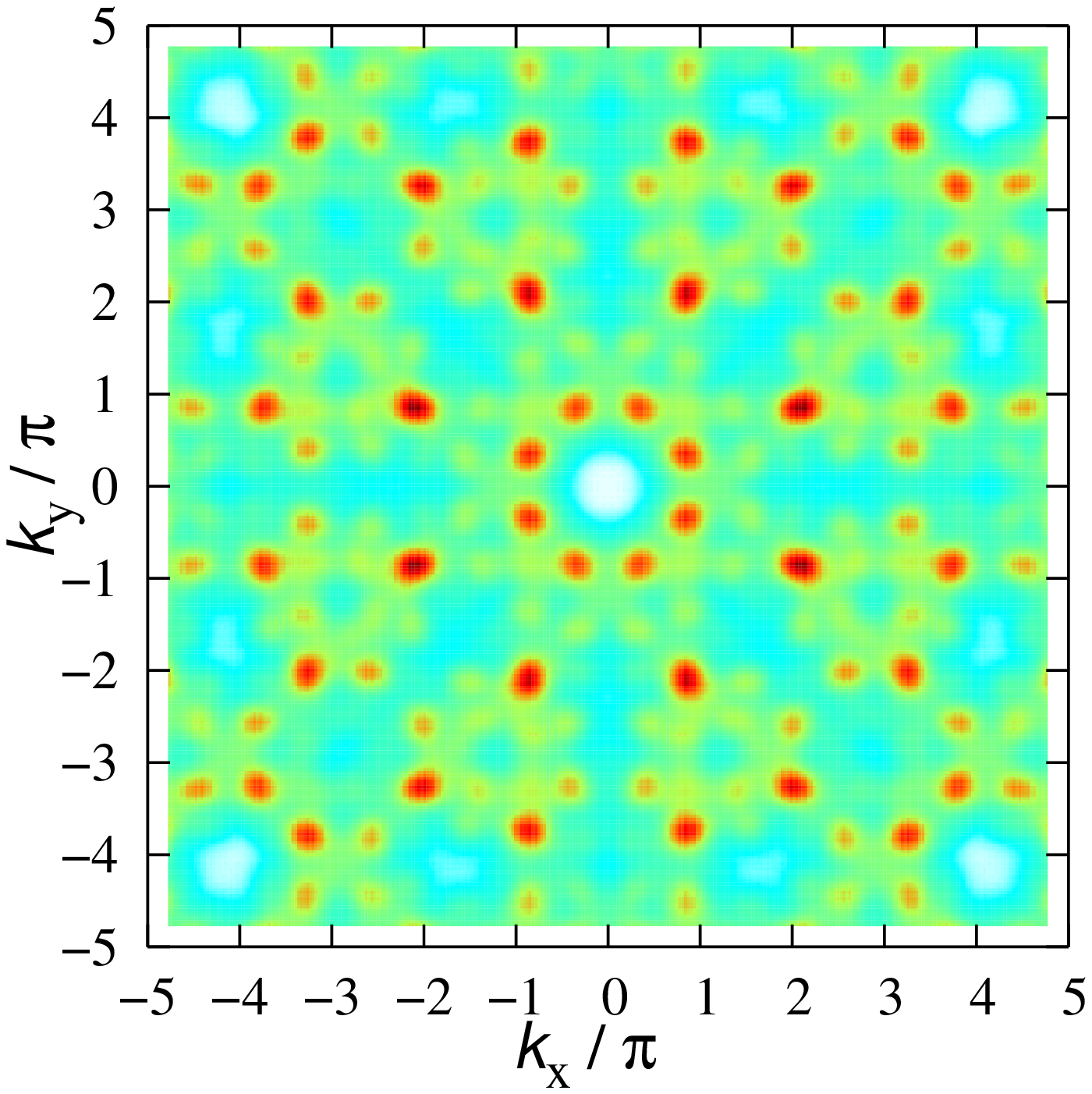}
\includegraphics[width=4cm]{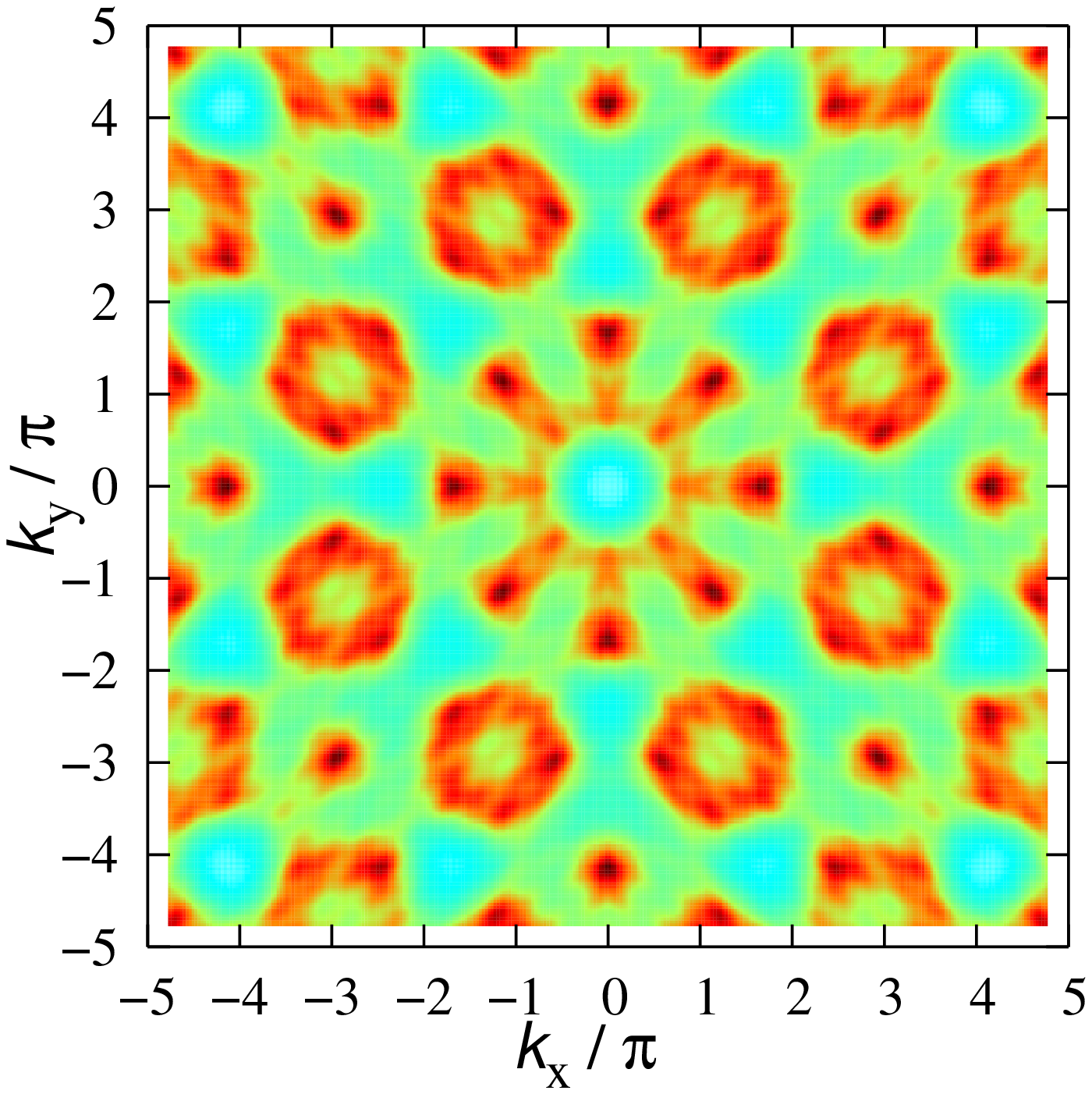}
\caption{
Intensity plot of the dynamical spin structure factor, $S({\mathbf k},\omega)$,
for fixed $\omega/JS=4.6$ (left) and $\omega/JS=6.4$ (right)
for the $S=1/2$ Heisenberg antiferromagnet on  the 1393 sites approximant of the 
octagonal tiling.
}
\label{fig:Sxx_o_3}
\end{center}
\end{figure}

\begin{figure}[t]
\begin{center}
\includegraphics[width=4cm]{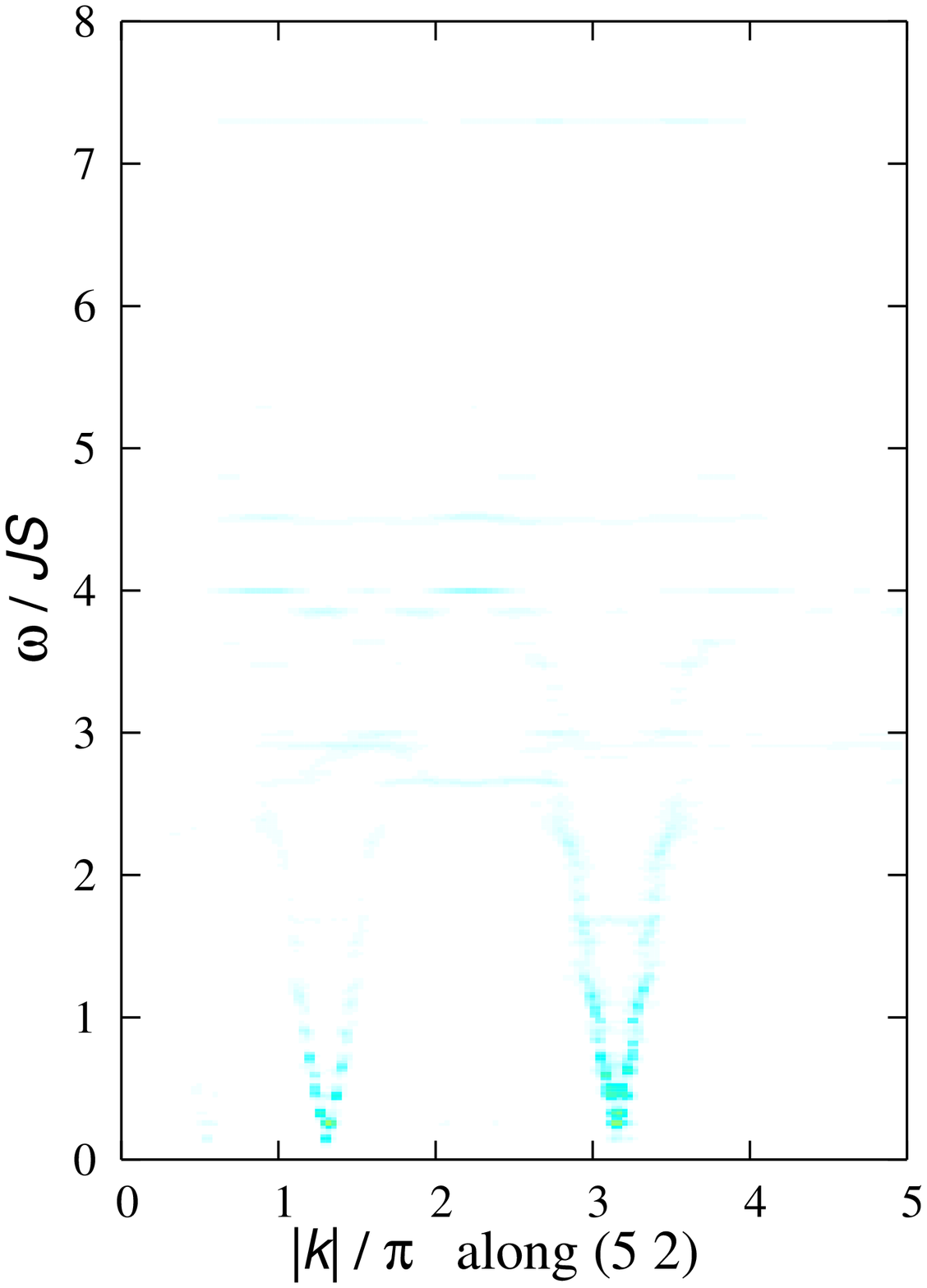}
\includegraphics[width=4cm]{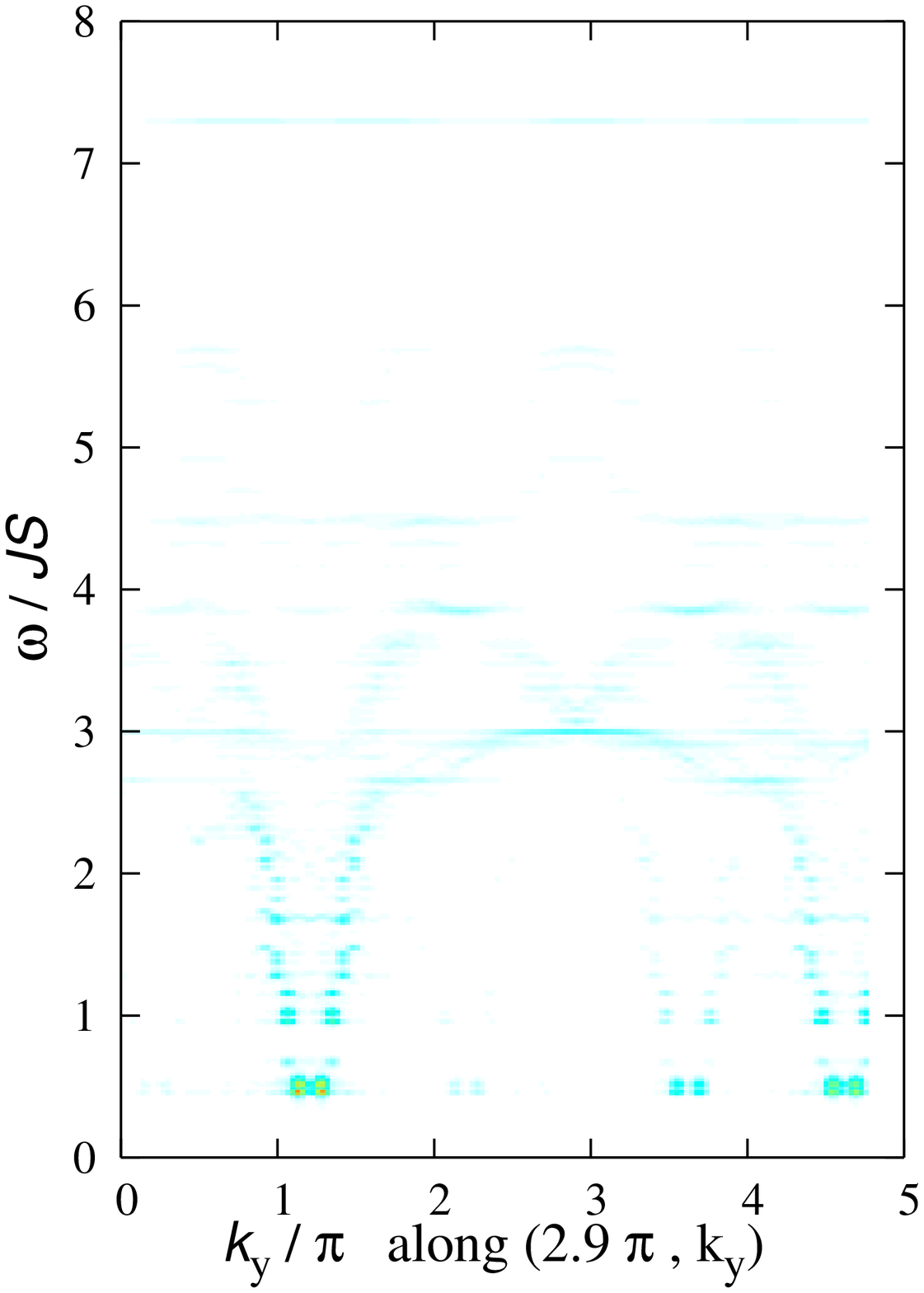}
\caption{
Intensity plot of the dynamical spin structure factor, $S({\mathbf k},\omega)$,
along the $(5,2)$ momentum space direction (left) and along the line $(2.9\pi,k_y)$ (right)
for the $S=1/2$ Heisenberg antiferromagnet on  the 1393 sites approximant of the 
octagonal tiling.
}
\label{fig:Sxx_k_1}
\end{center}
\end{figure}

\begin{figure}[t]
\begin{center}
\includegraphics[width=4cm]{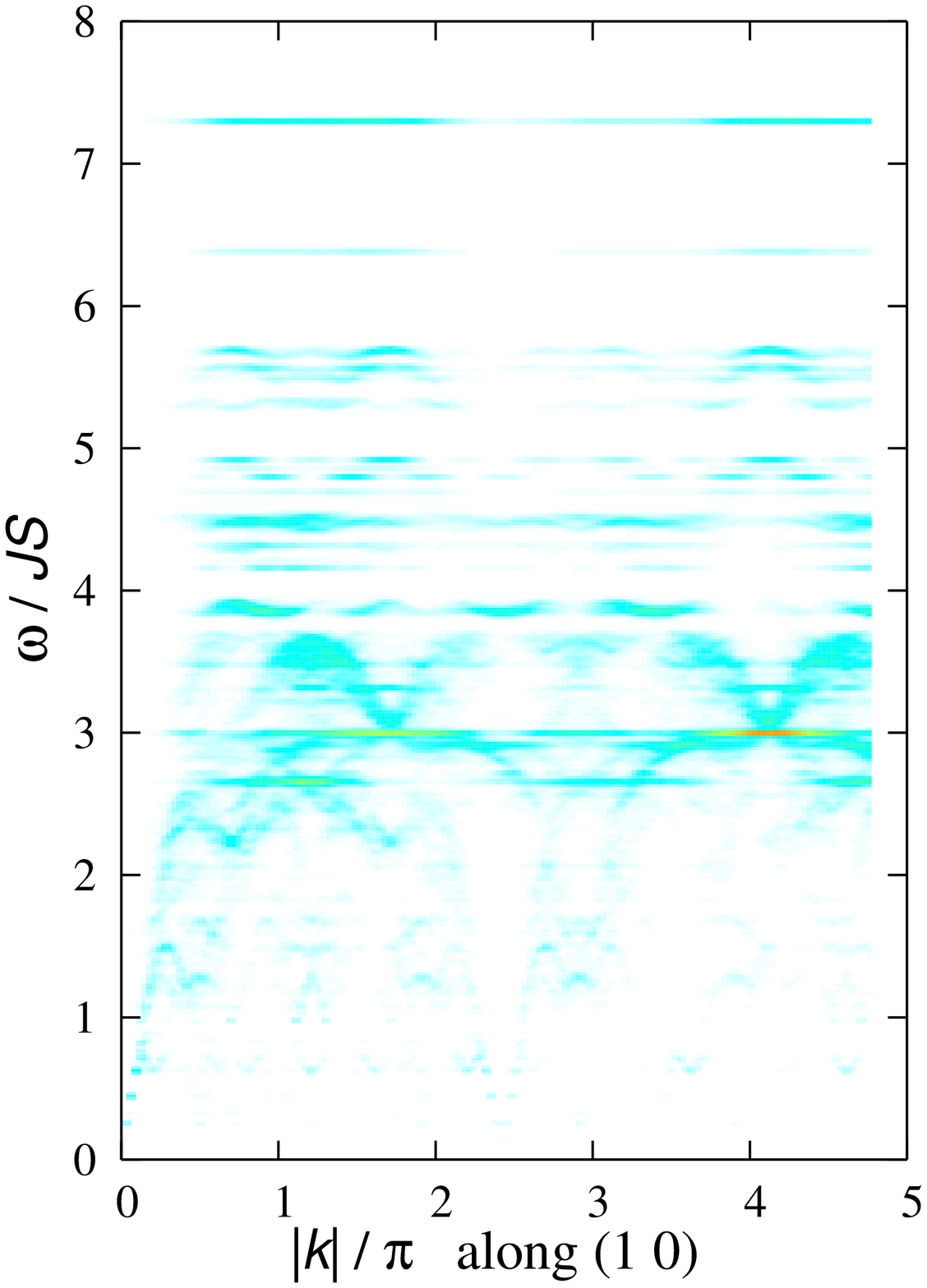}
\includegraphics[width=4cm]{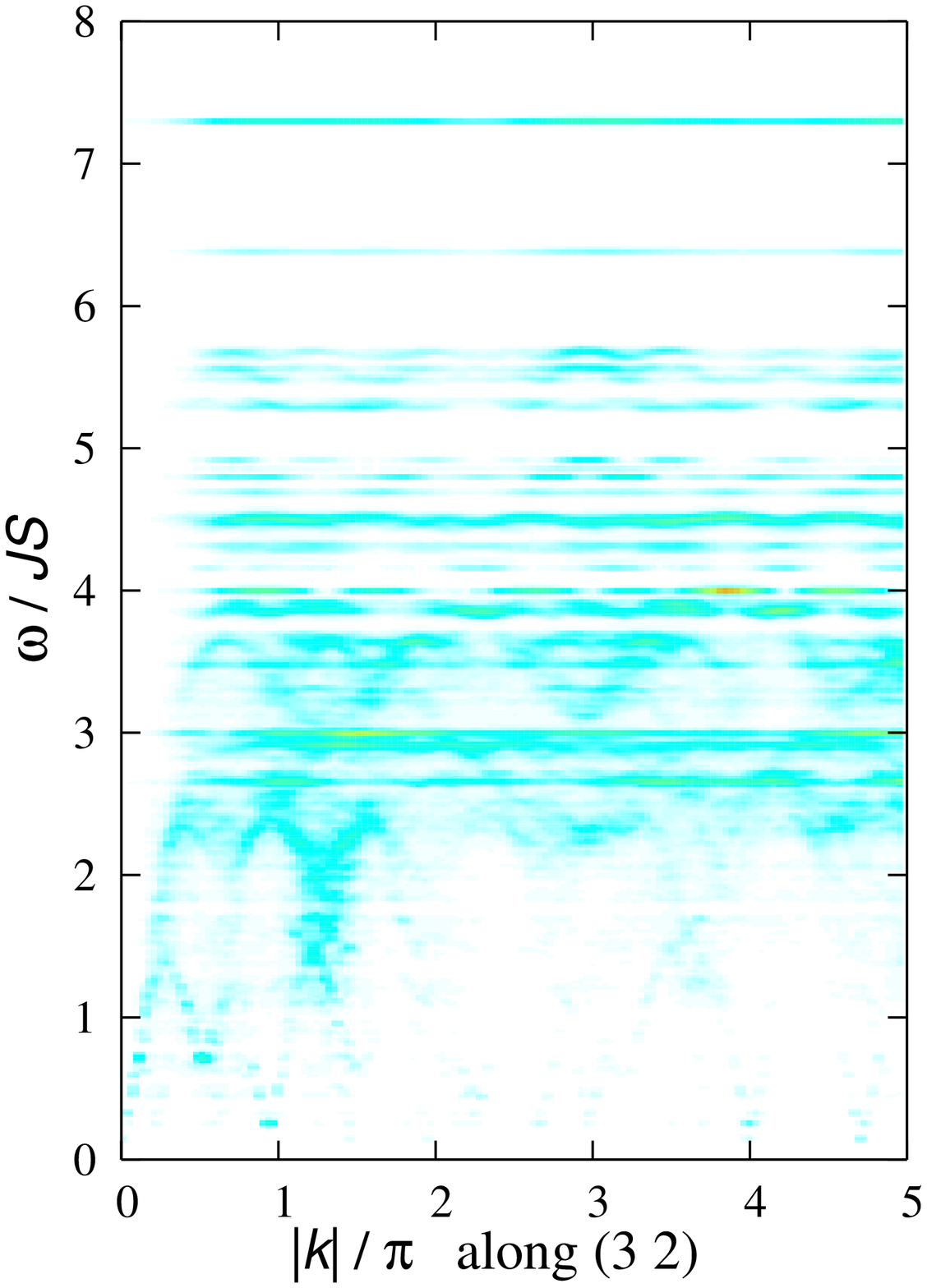}
\caption{
Intensity plot of the dynamical spin structure factor, $S({\mathbf k},\omega)$,
along the $(1,0)$  momentum space direction (left) and the $(3,2)$ direction (right)
for the $S=1/2$  Heisenberg antiferromagnet on  the 1393 sites approximant of the 
octagonal tiling.
}
\label{fig:Sxx_k_2}
\end{center}
\end{figure}

The dynamical spin structure factor, accessible experimentally by inelastic neutron scattering, reflects the time-dependent 
spin-spin 
correlation functions, transformed to  momentum space, 
\begin{equation}
 S^{\perp}({\mathbf k},\omega)=\frac{1}{N_s}\sum_{i,j=1}^{N_s} e^{i{\mathbf k} \cdot ({\mathbf r}_i - {\mathbf r}_j)} 
S^{\perp}(i,j,\omega),
\end{equation}
with 
\begin{equation}
 S^{\perp}(i,j,\omega)=\frac{1}{2}\int_{-\infty}^{\infty} dt\: e^{i\omega t} \langle S^+_i(t) S^-_j(0) + S^-_i(t) S^+_j(0) 
\rangle.
\end{equation}
Using the normal-mode expansion, Eq. (\ref{eq:tildeT}), we obtain the following expression for $S^{\perp}(i,j,\omega)$ within 
linear 
spin-wave theory,
\begin{equation}
 S^{\perp}(i,j,\omega)=S \sum_k \left( U_{ik} U^*_{jk} + V^*_{ik} V_{jk} \right) \delta(\omega-\omega_k),
\end{equation}
from which $S^{\perp}({\mathbf k},\omega)$ is readily obtained using a fast Fourier transformation. Most spectral 
weight in $S^{\perp}({\mathbf k},\omega)$ 
is located at momenta corresponding to magnetic Bragg peaks, as seen by comparing the integrated dynamical spin structure  
factor,
\begin{equation}
 S^{\perp}({\mathbf k})=\int \frac{d\omega}{2\pi} \: S^{\perp}({\mathbf k},\omega),
\end{equation}
shown in the right of Fig.~\ref{fig:Sxx_Szk} to the static longitudinal structure factor, $S^{\parallel}({\mathbf k})$, 
in the left of Fig.~\ref{fig:Sxx_Szk}.
For a detailed analysis of the dynamical spin structure factor, 
we consider both constant-frequency 
scans, shown in Figs.~\ref{fig:Sxx_o_1} to~\ref{fig:Sxx_o_3}, as well as scans along various momentum 
space 
directions, 
shown in Figs.~\ref{fig:Sxx_k_1} and~\ref{fig:Sxx_k_2}. To increase the contrast in these figures, we have rescaled 
the data to the maximum value in each plot, separately. 

In order to compare the relative spectral weight at different energies, we show in Fig.~\ref{fig:Sxx_o_all}
the momentum-integrated spectral function,
\begin{equation}
S^{\perp}(\omega)=\int\frac{d^2 k}{(2\pi)^2} \: S({\mathbf k},\omega).
\end{equation}
Compared to the DOS, 
we find that apart from the low-energy 
region below $\omega/JS\approx 2$, the shape of $S^{\perp}(\omega)$ closely reflects the structures in the DOS. In the 
low-energy region, we observe an disproportionately large contribution to $S^{\perp}(\omega)$, in contrast to the low DOS 
in this region. The difference between $S^{\perp}(\omega)$, and the DOS, $\rho(\omega)$, in linear spin-wave theory is obtained
using Eq. (\ref{eq:tildeT}) as 
\begin{equation}
S^{\perp}(\omega)-S\rho(\omega)=2 S \sum_{i,k} |V_{ik}|^2  \delta(\omega-\omega_k),
\end{equation}
and indicates, that the extra contributions to $S^{\perp}(\omega)$ at energies $\omega/JS<2$ are due to the spectral predominance
of low energy quantum fluctuations, $\delta m_s(\omega)$, Eq.~(\ref{eq:dms}), shown in Fig.~\ref{fig:dms}. 
In the following, we first consider the dynamical spin structure factor in this low-energy region, which 
is dominated by magnetic Bragg scattering. 

In the elastic limit of $S^{\perp}({\mathbf k},\omega\rightarrow 0)$, shown in the left of Fig.~\ref{fig:Sxx_o_1}, we 
can indeed identify the
magnetic Bragg peak positions of the static longitudinal structure factor, shown in Fig.~\ref{fig:Sxx_Szk}.
Furthermore, for all energies up to $\omega/JS\approx 2$, similar patterns as in $S^{\perp}({\mathbf k},0)$ are 
observed, albeit with the 
width of the peaks increasing upon increasing the energy. Eventually, these peaks  evolve into ring-like structures, 
centered around the magnetic Bragg peaks,  such 
as shown for $\omega/JS=1.8$ in the right part of Fig.~\ref{fig:Sxx_o_1}.  This is a clear indication for
magnetic soft-modes at low energies, which dominate the magnetic response in this energy regime.

Typical examples of $S^{\perp}({\mathbf k})$ at higher frequencies are shown in 
Fig.~\ref{fig:Sxx_o_2},
and~\ref{fig:Sxx_o_3}. We find all plots to exhibit an eightfold overall 
symmetry, as expected for the octagonal tiling. However, the positions of the dominant peaks are different from the 
magnetic Bragg peaks 
found below $\omega/JS\approx2$. For example, for $\omega/JS=3.0$ (right of Fig.~\ref{fig:Sxx_o_2}), most spectral weight is 
located at ${\mathbf k}\approx(0, 1.3 \pi)$, and
${\mathbf k}\approx(0, 4.3 \pi)$, as well as symmetry-related momenta. Peaks at these momenta are absent in both 
$S^{\parallel}({\mathbf k},0)$, and the nuclear Bragg scattering (Fig.~5 (a) of Ref.~\onlinecite{wessel}). In fact, we do 
not 
observe pronounced spectral weight at nuclear Bragg peak positions for any finite energy cut: to give a  further example
of this fact, we show 
$S^{\perp}({\mathbf k})$ at $\omega/JS=4.6$ in the left of Fig.~\ref{fig:Sxx_o_3}.
Upon changing the energy-level of the cut  only  slightly, the patterns found 
in the high-energy
region
change more drastically 
than those in the low-energy regime. As an example, in the left 
of Fig.~\ref{fig:Sxx_o_2}, we show $S^{\perp}({\mathbf k})$ for $\omega/JS=2.9$. 
Compared to $S^{\perp}({\mathbf k})$ at $\omega/JS=3.0$ (right of Fig.~\ref{fig:Sxx_o_2}), we indeed find a different set of 
dominant peaks. 

The patterns in $S^{\perp}({\mathbf k})$ for energies corresponding to the high-energy pockets in the DOS 
(c.f.~Fig.~\ref{fig:dos})
consist of more diffusive structures than those at lower energies.
For example, we find broad ring-like structures, centered around the magnetic Bragg peaks
in $S^{\perp}({\mathbf k})$ for $\omega/JS=6.4$, shown in the right of Fig.~\ref{fig:Sxx_o_3}.
The appearance of such diffusive structures at high energies is expected from the results of 
Sec.~\ref{sec:exci}, where the 
the high-energy excitations were 
found 
to be spatially less extended than the low-energy states. 
This reduced spatial extent leads to smeared 
diffraction patterns observed in Fig.~\ref{fig:Sxx_o_3}.

In order to analyze the momentum dependence of the dominant peaks in the dynamical spin structure factor, we choose 
representative 
directions in momentum 
space, and plot $S^{\perp}({\mathbf k},\omega)$ along such cuts. We first consider momenta lying along the $(5\; 2)$ 
direction, 
shown in the left of 
Fig.~\ref{fig:Sxx_k_1}. 
This cut passes through two of the major magnetic Bragg peaks, namely at $|{\mathbf k}_B|/\pi\approx 1.3$,  
and $3.2$. In the low-energy region, below $\omega/JS\approx 2$, most spectral weight is located along straight lines, emerging 
from the magnetic Bragg peaks, and  with spectral weight that increases for decreasing energy, characteristic of magnetic 
soft-modes.
Near the magnetic Bragg peaks, ${\mathbf k}_B$, we thus observe linear dispersion relations
of the magnetic excitations, $\omega=c |{\mathbf 
k}-{\mathbf k}_B|$, with 
an estimated spin-wave velocity $c/JS \approx 2.1$, which is of similar order of magnitude than the linear spin-wave 
result for the square lattice 
($c/JS=\sqrt{8}\approx 2.83$). Similar soft-modes are 
also observed along the cut of constant $k_x/\pi=2.9$, shown in the right of Fig.~\ref{fig:Sxx_k_1}, including a
further magnetic Bragg peak at ${\mathbf k}_B~\approx(2.9\pi,4.8\pi)$.
The emergence of these linear low-energy dispersion relations is consistent with the  linear 
low-energy DOS found in Sec.~\ref{sec:exci}, and furthermore explains the ring-like structures 
in $S^{\perp}({\mathbf k})$, as seen for $\omega/JS=1.8$ in the right of Fig.~\ref{fig:Sxx_o_2}. 

In spite of the absence of translational symmetry, the dynamical spin structure factor of 
the antiferromagnetic quasicrystal clearly exhibits 
the presence of soft-modes near the magnetic Bragg peak positions of the quasiperiodic 
crystal. We expect such a generic feature of magnetic long-range order to be present also in other magnetically ordered 
quasicrystals. 

In contrast, at high frequencies, $\omega/JS>5.5$, we do not observe any significant dispersion of the spectral weight 
distribution. Instead, we find two flat bands of only slightly modulated spectral weight located near $\omega/JS\approx 6.4$, 
and $7.3$ in Fig.~\ref{fig:Sxx_k_2}, which 
correspond
to the two isolated pockets of the spin-wave DOS in  Fig.~\ref{fig:dos}. We consider this  observation as further 
indication for the limited spatial extent of the corresponding eigenstates, concluded in Sec.~\ref{sec:exci} from the finite size 
scaling behavior of the inverse participation ratio.

In the intermediate energy regime, between  $\omega/JS\approx 2$, and $\omega/JS\approx 5.5$, the distribution of 
spectral weight is more complex, and can be accounted for by  band-like segments, which recur at different energies 
and 
with varying bandwidths, as shown in  Fig.~\ref{fig:Sxx_k_2}. 
At points of increased spectral weight, such as for
$\omega/JS=3$ near $|{\mathbf k}|/\pi\approx 1.8$, and $4.2$, in the left of Fig.~\ref{fig:Sxx_k_2}, 
the corresponding gapped modes furthermore show a linear dispersion relation.
In addition, we find bifurcations emerging as branches of these band-segments
extending towards low-energies. This is seen for example in the right part of Fig.~\ref{fig:Sxx_k_2}, for 
momenta $|{\mathbf k}|<\pi$. 
Such self-similar structures might have been expected to dominate the dynamical spin structure factor,  due to the geometric 
properties of the octagonal tiling, reflecting its inflation symmetry. 
Nevertheless, we find them well separated in energy from more conventional low-energy features, that reflect the 
magnetic order in this system. 

\subsection{Local Dynamical Spin Susceptibility}
\label{sec:loca}
While the dynamical spin structure factor thus exhibits the peculiar nature of the excitations in the self-similar 
quasiperiodic system, the different local environments of the magnetic moments in the quasicrystal are accessible from the 
local dynamical spin 
susceptibility, the imaginary part of which  at 
each lattice site $i$ is given by
\begin{equation}
 \chi''_{\mathrm local}(i,\omega)=S^{\perp}(i,i,\omega)=S \sum_k \left( |U_{ik}|^2 + |V_{ik}|^2 \right) \delta(\omega-\omega_k),
\end{equation}
within linear spin-wave theory. The local dynamical spin
susceptibility
is accessible in nuclear magnetic resonance experiments,
in the form of Knight-shifts at nuclear sites in the vicinity of the magnetic sites. Here, we study the properties
of $\chi''_{\mathrm local}$ for the Heisenberg model on the octagonal tiling, as an example of a quasiperiodic lattice structure.

\begin{figure}[t]
\begin{center}
\includegraphics[width=8cm]{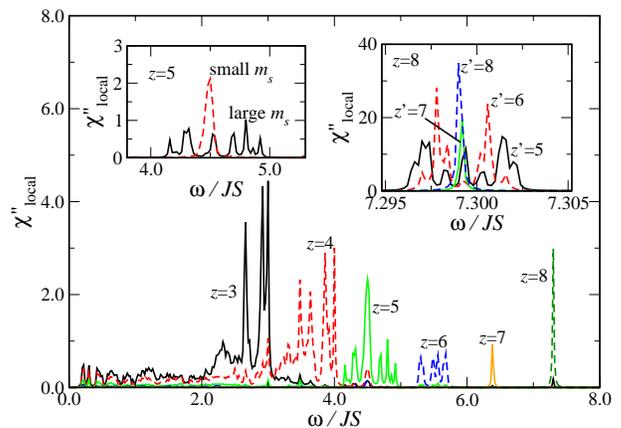}
\caption{
Imaginary part of the local dynamical spin susceptibility, $\chi''_{\mathrm local}(i,\omega)$, 
for the $S=1/2$ Heisenberg antiferromagnet on the  octagonal tiling,
averaged separately over sites with
coordination numbers $z=3$ to $8$.
The left inset shows $\chi''_{\mathrm local}(i,\omega)$ for the fivefold sites, averaged separately over sites with a
small and  large staggered magnetization, respectively. The right inset shows $\chi''_{\mathrm local}(i,\omega)$ for the eightfold sites,
averaged separately over sites with different behavior under deflation transformation, grouped according to the value of $z'$.
}
\label{fig:sii}
\end{center}
\end{figure}

In Fig.~\ref{fig:sii}, $\chi''_{\mathrm local}$ is shown for the largest approximant, averaged separately for 
sites of different coordination. We observe a broad spread in the signal, with characteristic energies that
increase linearly from $\omega/JS\approx2.5$ for threefold sites to $\omega/JS\approx7.3$ for eightfold sites. 
Furthermore,
the energy-range over which there is a large signal narrows for sites of increasing coordination, reflecting a similar trend in the 
spread of the local staggered moments (c.f. Fig.~\ref{fig:lsm}). The inequivalent local environments of the various sites are responsible for this
extended range of signals. For example, the left insets of Fig.~\ref{fig:sii} exhibits that the two types of fivefold sites
show a different frequency dependence of the dynamical spin susceptibility. Namely, sites with a smaller moment have signals inside a narrow region 
near 
$\omega/JS\approx 4.5$, whereas sites with a larger moment produce signals over a more extended region, ranging from $\omega/JS\approx4.2$ to 
$\omega/JS\approx 5$. The high symmetry eightfold sites also exhibit a characteristic splitting in the local spin susceptibility, as seen in the 
right inset of Fig.~\ref{fig:lsm}. Here, the individual signals are labeled by the value of $z'$
for the site from which this signal results.
We observe from Fig.~\ref{fig:lsm}, that the widths of the signals narrow towards  $\omega/JS\approx 7.3$ for increasing values of $z'$.
The local dynamical spin susceptibility thus reflects the hierarchical structure of the local moment distribution of the eightfold sites. 
We expect such features to be generic properties of quasiperiodic magnets, eventually seen  
in nuclear magnetic resonance experiments on real quasiperiodic magnetic systems.

\section{Conclusions}
\label{sec:conc}
We studied the antiferromagnetic spin-1/2 Heisenberg model on the octagonal tiling, a two-dimensional 
quasiperiodic lattice structure, using linear spin-wave theory in a real space formulation. This approach was found to 
quantitatively reproduce previous quantum Monte Carlo results on static magnetic ground state properties of this system.
The spin-wave excitation spectrum was found to consist of  magnon-like low-energy soft-modes with a linear 
dispersion relation near the magnetic Bragg peaks, characteristic to long-range magnetic order.
It will be interesting to confirm the existence of  such linear soft-modes in the octagonal tiling
in future quantum Monte Carlo studies of the dynamical spin structure factor.

In addition, the dynamical spin structure factor shows
self-similar structures and bifurcations,
as well as flat bands
at higher energies.
We expect such features to be 
generic to magnetic quasicrystals, which might
eventually become observable in neutron scattering 
experiments on magnetically ordered quasicrystals. 

Within the spin-wave approach, it is possible to include magnetic frustration, offering the 
possibility of modeling
more realistic quasiperiodic lattice structures, and their influence on magnetic properties. 
Starting in 
the unfrustrated limit, and increasing the magnetic frustration, 
the evolution  of the classical N\'eel state can be examined, as well as the potential  relevance of the
multifractal excitations for its breakdown in the case of strong frustration. 
Another route to magnetic disorder, which can be taken at least theoretically, is by means of a quasiperiodic bilayer, which 
is expected to 
show a quantum phase transition upon increasing the interlayer coupling, due to local singlet formation. The presence of 
multifractal excitations in the quasicrystal might be of possible relevance to quantum criticality in the transition 
between the N\'eel ordered state and the disordered, gaped state. We leave such studies for future research.

 \section*{Acknowledgments}

We thank P. Frigeri, S. Haas, O. Nohadani, D. Rau, M. Sigrist, and
S. Wehrli for fruitful discussions. The support of the MaNEP project of Swiss National Science Foundation is gratefully acknowledged.  Parts of the numerical calculations were done
using the ALPS project library~\cite{alps}, and performed
on the Asgard Beowulf cluster and Superdome at ETH Z\"urich.

\appendix

\section{Numerical Bogoliubov Transformation}

In this appendix we describe an alternative numerical scheme of finding the Bogoliubov
transformation of the spin-wave Hamiltonian\cite{avery}. We need to construct a  matrix $T$, which diagonalizes the Hamiltonian
matrix $M$ defined in Eq.~(\ref{eq:Mijkl}), and also satisfies the
constraint in Eq.~(\ref{eq:constr}). We therefore need to simultaneously solve 
\begin{equation}
  T^\dag M T = \Omega,\quad \text{and} \quad   T^\dag \Sigma T = \Gamma.
\end{equation}

The matrix $T$ can be constructed in two steps as follows:
In a first step, an eigenvector matrix $Z$ and the set of eigenvalues
$\lambda_i$ is obtained for the non-Hermitian eigenvalue problem~\cite{lapack},
\begin{equation}
  \Sigma M Z  = Z \Lambda,
\end{equation}
where $\Lambda = \text{diag}(\lambda_1, \ldots, \lambda_{N_n})$. 
The columns of $Z$ are the right eigenvectors of $\Sigma M$.
Hermitian conjugation of
the above equation yields
\begin{equation}
  (\Sigma Z^\dag \Sigma) \Sigma M = \Lambda ( \Sigma Z^\dag \Sigma ),
\end{equation}
so that the rows of $\Sigma Z^\dag \Sigma$ form the left eigenvectors of $\Sigma M$.

In a second step, we diagonalize the Hermitian matrix,
\begin{equation}
  L = Z^\dag \Sigma Z,
\end{equation} 
obtaining a unitary matrix $U$, such that
\begin{equation}
  U^\dag L U = \text{diag}(l_1, \ldots, l_{N_n}).
\end{equation}
Furthermore, from the eigenvalues $l_i$ of $L$ we construct 
the diagonal matrix $l^{-1/2}$, defined as
\begin{equation}
  l^{-1/2}_{ij} = 
    \delta_{ij} |l_i|^{-1/2}.
\end{equation}
In case that the eigenvalues of $\Sigma M$ are non-degenerate,
performing this second step is trivial: In this case $l_i$ is the
$\Sigma$-norm of the corresponding eigenvector, and $U = 1$.
We finally obtain the transformation $T$ as:
\begin{equation}
  T = Z U l^{-1/2}.
\end{equation}
Indeed, since 
\begin{equation}
\label{eq:app_L}
 \bar \Gamma = T^\dag \Sigma T
\end{equation}
satisfies $\bar\Gamma^2 = 1$ by construction,
$T$ satisfies the constraint Eq.~(\ref{eq:constr}). 
Furthermore, right and left eigenvectors belonging to different
eigenvalues are orthogonal. Thus $L$ is block diagonal with blocks
corresponding to degenerate subspaces of $\Sigma M$. It follows that
$U$ is block diagonal as well. Therefore $U$ only mixes columns of $Z$
belonging to the same eigenvalue of $\Sigma M$, and consequently $T$
satisfies
\begin{equation}
  \Sigma M T = T \bar \Lambda.
\end{equation}
Multiplying the above equation from the left by $T^{-1} = \Gamma T^\dag \Sigma$ 
(which follows from Eq.~(\ref{eq:app_L})), yields
\begin{equation}
  T^\dag M T = \Lambda \bar \Omega.
\end{equation}
Hence, the matrix $T$  also diagonalizes the Hamiltonian matrix $M$.
The energies of the bosonic eigenmodes are given as $\omega_i=JS\Gamma_{ii}\lambda_i$, respectively.


\addcontentsline{toc}{section}{References}


\begin{thebibliography}{8.}

\bibitem{manousakis} E. Manousakis, Rev. Mod. Phys. {\bf 63}, 1 (1991), and references therein.
\bibitem{plaquette} A. Koga, S. Kumada, N. Kawakami, J. Phys. Soc. Jpn. {\bf 68} (7), 2373 (1999).
\bibitem{laeuchli} A. L\"auchli, S. Wessel, and M. Sigrist,  Phys. Rev. B {\bf 66}, 014401 (2002).
\bibitem{misguich} Gregoire Misguich, Claire Lhuillier, Report cond-mat/0310405, and references therein.
\bibitem{depletion} A. Sandvik, Phys. Rev. B {\bf 66}, 024418 (2002).
\bibitem{mucciolo} E. R. Mucciolo, A. H. Castro Neto, and C. Chamon, Report cond-mat/0402102.
\bibitem{sato} T. J. Sato, H. Takakura, A. P. Tsai, K. Shibata, K. Ohoyama, and K. H. Andersen, Phys. Rev. B {\bf 61}, 476 
(2000).
\bibitem{cdmgtb} T. J. Sato, H. Takakura, G. Guo, A. P. Tsai, and K. Ohoyama, J. of Alloys and Compounds, {\bf 342}, 365 
(2002).
\bibitem{absence} T. J. Sato, H. Takakura, A. P. Tsai, and K. Shibata, Phys. Rev. Lett. {\bf 81}, 2364 (1998).
\bibitem{wessel} S. Wessel, A. Jagannathan, S. Haas, Phys. Rev. Lett. {\bf 90}, 177205 (2003).
\bibitem{anu} A. Jagannathan, Phys. Rev. Lett. {\bf 92}, 047202 (2004).
\bibitem{lifshitz} R. Lifshitz and S. Even-Dar Mandel, Acta Cryst. A {\bf 60}, 167 (2004).
\bibitem{ron} Ron Lifshitz, Material Science and Engineering A {\bf 294}, 508 (2000).
\bibitem{vedmedenko} E. Y. Vedmedenko, U. Grimm, and R. Wiesendanger, Report cond-mat/0406373.
\bibitem{grimm}U. Grimm, and M. Schreiber, in {\it  Quasicrystals - Structure and Physical Properties}, ed. H.-R. Trebin 
(Wiley-VCH, Weinheim, 2003), and references therein.
\bibitem{levine} D. Levine and P. J. Steinhardt, Phys. Rev. B {\bf 34}, 596 (1986).
\bibitem{white} R. M. White, M. Sparks, and I. Ortenburger, Phys. Rev. {\bf 139}, 450 (1965).
\bibitem{duneau} M. Duneau {\it et al.}, J. Phys. A {\bf 22}, 4549 (1989).
\bibitem{schulz} A. Jagannathan and H. J. Schulz, Phys. Rev. B {\bf 55}, 8045 (1997).
\bibitem{spinwave} K. Yosida, {\it Theory of Magnetism} (Springer, 1996).
\bibitem{blaizot} J.-P. Blaizot and G. Ripka, {\it Quantum Theory of Finite Systems} (MIT Press, Cambridge, MA, 1986).
\bibitem{lapack} Using standard numerical routines, for example from LAPACK.
\bibitem{lang} S. Lang, {\it Algebra}, (3rd Edition, Addison-Wesley 1993).
\bibitem{einarsson} T. Einarsson and H. J. Schulz, Phys. Rev. B {\bf 51}, 6151 (1995).
\bibitem{wessel_e0} S. Wessel, unpublished.
\bibitem{footnote1} We verified, that the shown averages are indeed representative of the individual contributions.
\bibitem{anu2000} A. Jagannathan, Phys. Rev. B. {\bf 61}, R834 (2000).
\bibitem{wegner} F. Wegner, Z. Phys. B {\bf 36}, 209 (1980).
\bibitem{schreiber} M. Schreiber and H. Gurssbach, Phys. Rev. Lett. {\bf 67}, 607 (1991).
\bibitem{alps} M. Troyer {\it et al.}, Lecture Notes in Computer
    Science {\bf 1505}, 191 (1998). Source codes of the library can
    be obtained from \verb|http://alps.comp-phys.org/| .
\bibitem{avery} Avery J. {\it Creation and annihilation operators} (McGraw-Hill, 1974), appendix 1 and references therein.
\end{thebibliography}
\end{document}